\documentclass[aps,prx,floatfix,twocolumn,superscriptaddress]{revtex4-2}

\usepackage{mathrsfs}
\usepackage{graphicx}
\usepackage[english]{babel}
\usepackage{amsmath}
\usepackage{amssymb}
\usepackage{xcolor}
\usepackage{cancel}
\usepackage[normalem]{ulem}

\usepackage{relsize}
\usepackage{CJK}
\usepackage{dsfont}
\usepackage{bm}

\newcommand{\me}{\mathrm{e}}
\newcommand{\mi}{\mathrm{i}}

\newcommand{\dif}{\mathrm{d}}

\newcommand{\bk}{\mathbf{k}}

\newcommand{\bB}{\mathbf{B}}

\newcommand\px{\mathrel{/\mkern-5mu/}}
\allowdisplaybreaks
\begin{document}

\title{Local geometry and quantum geometric tensor of mixed states}

\author{Xu-Yang Hou}
\affiliation{School of Physics, Southeast University, Jiulonghu Campus, Nanjing 211189, China}

\author{Zheng Zhou}
\affiliation{School of Physics, Southeast University, Jiulonghu Campus, Nanjing 211189, China}
\author{Xin Wang}
\affiliation{School of Physics, Southeast University, Jiulonghu Campus, Nanjing 211189, China}

\author{Hao Guo}
\email{guohao.ph@seu.edu.cn}
\affiliation{School of Physics, Southeast University, Jiulonghu Campus, Nanjing 211189, China}
\affiliation{Hefei National Laboratory, Hefei 230088, China}

\author{Chih-Chun Chien}
\email{cchien5@ucmerced.edu}
\affiliation{Department of physics, University of California, Merced, CA 95343, USA}

\begin{abstract}
The quantum geometric tensor (QGT) is a fundamental concept for characterizing the local geometry of quantum states. After casting the geometry of pure quantum states and extracting the QGT, we generalize the geometry to mixed quantum states via the density matrix and its purification. The gauge-invariant QGT of mixed states is derived, whose real and imaginary parts are the Bures metric and the Uhlmann form, respectively. In contrast to the imaginary part of the pure-state QGT that is proportional to the Berry curvature, the Uhlmann form vanishes identically for ordinary physical processes. Moreover, there exists a Pythagorean-like equation that links different local distances and reflect the underlying fibration. The Bures metric of mixed states is shown to reduce to the corresponding Fubini-Study metric of the ground states as temperature approaches zero, establishing a  correspondence despite the different underlying fibrations. We also present two examples with contrasting local geometries and discuss experimental implications.
\end{abstract}

\maketitle

\section{Introduction}
The geometry of quantum states has sparked lasting research interest~\cite{Nakahara,Bengtsson_book} and played an indispensable role in quantum statistical mechanics, quantum information, condensed matter, and atomic, molecular, and optical physics \cite{IG_Book,Bohm03,KOLODRUBETZ20171,QGTCMP80,cmp/1103904831,RevModPhys.82.1959,PhysRevB.74.085308,PhysRevLett.72.3439}. Unlike the topology of quantum systems that reflects global properties of the underlying structures via quantized indices~\cite{ZhangSCRMP,KaneRMP,ChiuRMP}, the geometry of quantum states can be sensitive to local intricacies. The quantum geometric tensor (QGT) is a fundamental concept for characterizing the variation between quantum states \cite{QGTCMP80,BRODY200119,QGT10,KOLODRUBETZ20171}. For pure states, the QGT is a complex quantity, whose real part gives the Fubini-Study metric \cite{EGUCHI1980213} that assesses the local distance between physically distinct quantum states while the imaginary part is proportional to the Berry curvature \cite{Simon83,Berry84} and related to the topology.

Pure quantum states are considered equivalent if they differ by a global phase~\cite{MQM}.
Taking into account the equivalence, the phase space of pure states can be identified~\cite{Bengtsson_book}.
The QGT is then a complex metric defined on the phase space, which measures the local geometry and is invariant under  gauge  transformation.
There have been many studies on the QGT of pure states, including proposals to extract the QGT  \cite{PhysRevB.87.245103,PhysRevB.97.201117,PhysRevB.97.041108,PhysRevLett.121.020401,PhysRevB.97.195422,PhysRevLett.124.197002}, experimental
realizations using photoluminescence of exciton-photon polaritons~\cite{Gianfrate2020}, Rabi oscillation of an NV center in diamond~\cite{10.1093/nsr/nwz193}, quench or periodic driving of a superconducting qubit~\cite{PhysRevLett.122.210401}, Bloch state tomography of cold atoms~\cite{Yi23}, and transmission measurements of plasmonic lattices~\cite{Cuerda23}, as well as theoretical works on optical response~\cite{Ahn22}, quantum fluctuations of the QGT~\cite{PhysRevLett.130.036202}, quantum phase transitions \cite{PhysRevLett.99.100603}, superfluidity in flat bands \cite{PhysRevLett.117.045303}, generalizations to $N$-band systems \cite{PhysRevB.104.085114}, topological semimetals~\cite{PhysRevX.10.041041} or other topological matter \cite{PhysRevB.90.165139,PhysRevA.92.063627,PhysRevB.105.045144,PhysRevLett.121.170401}, and PT-symmetric systems~\cite{PhysRevA.99.042104}. By considering the QGT as the second cumulant of the fidelity susceptibility, Ref.~\cite{PhysRevA.108.032218} studied other relevant quantities.

While most of the previous studies are about pure states, we move forward to formulating the QGT of mixed quantum states.
The equivalence among mixed states due to the phase factors and the phase space of mixed states will be be identified before we derive the gauge-invariant QGT for mixed states and unveil rich geometric properties behind the QGT. To present a complete picture of the QGT, we first examine the local geometry of the phase space of pure states, which is a K$\ddot{\text{a}}$hler manifold~\cite{Nakahara}, and its fibration that gives rise to a Pythagorean-like equation among the distances in different spaces.

For mixed quantum states, we will identify the phase space
and fibration by taking hints from the pure-state results.
While the QGT of pure states can be obtained through gauge-invariant modifications to the `raw' metric, we will show that a similar method applies to mixed states via Uhlmann's approach of the topology of full-rank density matrices~\cite{Uhlmann86, OurDPUP, 10.21468/SciPostPhysCore.6.1.024}. We find that all gauge-invariant modifications to the real part of the raw metric of mixed states reduce to the Bures metric within this framework, and a similar modification to the imaginary part causes it to vanish identically.
To our best knowledge,
the phase space of mixed states is not a K$\ddot{\text{a}}$hler manifold. Thus, the imaginary part of the QGT is not proportional to the Uhlmann curvature of mixed states, in contrast to the pure-state case.

The generalization of the QGT to mixed states serves another illustration of the subtle similarities and differences between pure and mixed states. While the Uhlmann bundle of full-rank density matrices is a trivial bundle~\cite{DiehlPRB15} and causes the associated characteristic classes to vanish, the Uhlmann holonomy and the Uhlmann phase can be nontrivial
and exhibit the Uhlmann-Berry correspondence as temperature approaches zero~\cite{OurUB}. Similarly, we will show that the mixed-state phase space lacks the K$\ddot{\text{a}}$hler structure but still exhibits an analogous Pythagorean-like relation between the raw metric and the metric on the phase space. Although the fibrations of pure and mixed states are different, we present a proof that the Bures metric can reduce to its pure-state counterpart, the Fubini-Study metric, as the temperature approaches zero and give rise to a correspondence between mixed and pure states.
We also compare the Bures distance and the Sj$\ddot{\text{o}}$qvist distance~\cite{PhysRevResearch.2.013344} by showing their different gauge conditions in a unified framework.
Finally, the uniqueness of the Bures distance within the framework based on the Uhlmann bundle explains why it is common in the literature on mixed states, despite many different arguments or derivations~\cite{PhysRevLett.72.3439,Wiseman_book,Bengtsson_book}.
Therefore, our work lays the foundation for systematic investigations of the geometry characterized by the QGT of mixed states.

Our main findings are: (1) The Uhlmann bundle provides a natural and solid foundation for deriving the QGT of mixed states with full-rank density matrices. (2) The gauge-invariant real part of the QGT reproduces the Bures metric, which approaches the Fubini-Study metric of pure states as temperature goes to zero. (3) The gauge-invariant imaginary part of the QGT is the Uhlmann form, which vanishes for regular systems in contrast to the Berry curvature of pure states. (4) A Pythagorean-like relation between the quantum distances of the total space and quantum phase space for mixed states emerges despite the lack of a K$\ddot{\text{a}}$hler structure like that of pure states.

The rest of the paper is organized as follows.
In Section \ref{SecII}, an overview of the QGT of pure states is presented,
and the geometric origins
of the real and imaginary components
are elucidated. In Section \ref{SecIII}, the QGT of mixed states from a geometric perspective is formulated, including the Pythagorean-like equation between different local distances and the gauge-invariant QGT. The relations to the Bures distance and Sj$\ddot{\text{o}}$qvist distance are explained and the vanishing of the imaginary part is derived.
We prove in Sec.~\ref{Sec4} that the Bures metric approaches the Fubini-Study metric as temperature goes to zero.
In Sec.~\ref{Sec5}, two examples of the QGT of mixed states are provided.
We also summarize some implications of the QGT of mixed states and challenges of going beyond the Uhlmann fibration. Sec.~\ref{sec:Con} concludes our work. Some details and derivations are summarized in the Appendix.

\section{QGT of pure states (Review) }\label{SecII}

\subsection{The quantum phase space}
Hereafter, the dimension of the quantum system of interest is assumed to be $N$. To contrast the similarities and differences between the geometries of pure and mixed states, we follow Refs.~\cite{Nakahara,Bengtsson_book} to review the physical space of the state-vectors. Readers familiar with the local geometry of pure states may proceed to Sec.~\ref{SecIII} directly.
In general, the linear space spanned by $N$-dimensional vectors is the $N$-dimensional complex vector space $\mathcal{H}=\mathbb{C}^N$. Since a state-vector $|\psi\rangle$ is physically equivalent to $c|\psi\rangle$ if $c$ is a nonzero complex number ($c\in \mathbb{C}^*=\mathbb{C}-\{0\}$), the phase space of pure states is a $(N-1)$-dimensional complex projective space, $P(\mathcal{H})=\mathbb{C}^N/\mathbb{C}^*\cong CP^{N-1}$, where the elements are complex rays.
A canonical projection connecting these two spaces is $\Pi:\mathcal{H}\rightarrow P(\mathcal{H})$ by collapsing any nonzero complex number $c$ \cite{GPbook}:  $\Pi(c|\psi\rangle)=|\psi\rangle$.
An alternative way to construct $CP^{N-1}$ begins with normalizing the state-vectors and obtaining a $(2N-1)$-dimensional real sphere $S(\mathcal{H}):=\{|\psi\rangle\in \mathcal{H}|\langle\psi|\psi\rangle=1\}\sim S^{2N-1}$. Since two points in $S(\mathcal{H})$ are physically equivalent $|\phi\rangle\sim|\psi\rangle$ if they differ only by a phase factor $|\psi\rangle=\me^{\mi\chi}|\phi\rangle$, $P(\mathcal{H})$ is the quotient space $S(\mathcal{H})/\text{U(1)}=CP^{N-1}$. The Hopf fibration \cite{Nakahara} corresponds to the $N=2$ case with $S^3/S^1=S^2$ since U(1)$\sim S^1$ and $CP^1\sim S^2$.
For a better distinction, we will use the tilde symbol to label a state $|\tilde{\psi}\rangle$ in $\mathcal{H}=\mathbb{C}^N$ or $S(\mathcal{H})$, and reserve the state $|\psi\rangle$ to indicate one in $P(\mathcal{H})=CP^{N-1}$.

A well-defined local distance between quantum states should be independent of any U(1) phase factor that only gives rise to an equivalence relation between the states. This indicates that the distance and the associated metric defined in the quantum phase space $CP^{N-1}$ must be invariant under arbitrary U(1) gauge transformations.
Furthermore, $CP^{N-1}$ is a K$\ddot{\text{a}}$hler manifold \cite{Nakahara} endowed with a K$\ddot{\text{a}}$hler metric, the real part of which is a Riemannian metric satisfying the gauge-invariant requirement while the imaginary part is a symplectic form proportional to the Berry curvature as a 2-form.

\subsection{The quantum distance}
Throughout the paper, we set $\hbar=k_B=1$ and apply the Einstein summation convention. Consider a family of normalized quantum states $|\tilde{\psi}(\mathbf{R})\rangle\in S(\mathcal{H})$, which depends on a set of real parameters $\mathbf{R}=(R^1,R^2,\cdots,R^k)^T\in \mathcal{M}$. Here $\mathcal{M}$ is a parameter manifold, whose dimension may not be equal to $N$. Examples of such a parameter manifold include the direction of an external magnetic field of a spin system or the Brillouin Zone of a lattice system. Following the idea of Refs.~\cite{QGTCMP80,QGT10}, the local distance between quantum states upon a variation of $\mathbf{R}$ is given by
\begin{align}\label{qd1}
\dif s^2=\big||\tilde{\psi}(\mathbf{R}+\dif \mathbf{R})\rangle-|\tilde{\psi}(\mathbf{R})\rangle\big|^2=\langle\partial_i\tilde{\psi}|\partial_j\tilde{\psi}\rangle\dif R^i\dif R^j.
\end{align}
It is tempting to choose the metric as $g_{ij}=\langle\partial_i\tilde{\psi}|\partial_j\tilde{\psi}\rangle$. However, this is not invariant under a local U(1) gauge transformation given by $|\tilde{\psi}(\mathbf{R})\rangle\rightarrow \me^{\mi\chi(\mathbf{R})}|\tilde{\psi}(\mathbf{R})\rangle$, so the quantum distance between physically equivalent states may not vanish. Interestingly, the imaginary part of $\langle\partial_i\tilde{\psi}|\partial_j\tilde{\psi}\rangle$ is invariant under a U(1) gauge transformation.

By plugging $\me^{\mi\chi}|\tilde{\psi}\rangle$ instead of $|\tilde{\psi}\rangle$ into Eq.~(\ref{qd1}), the quantum distance becomes
\begin{align}
&\dif s^2\rightarrow \dif s'^2\notag\\=&\left(\langle\partial_i\tilde{\psi}|\partial_j\tilde{\psi}\rangle-\mi \omega_i\partial_j\chi-\mi \omega_j\partial_i\chi+\partial_i\chi\partial_j\chi\right)\dif R^i\dif R^j,\notag
\end{align}
where \begin{align}\label{BA1}\omega_i=\langle\tilde{\psi}|\partial_i\tilde{\psi}\rangle=-\langle\partial_i\tilde{\psi}|\tilde{\psi}\rangle\end{align} is the Berry connection on $S(\mathcal{H})$. Under the same U(1) gauge transformation, it changes as $\omega_i\rightarrow \omega_i'=\omega_i+\mi\partial_i\chi.$ Thus, we can redefine the quantum distance as \cite{QGTCMP80,QGT10}
\begin{align}\label{qd3}
\dif s^2
&=\left(\langle\partial_i\tilde{\psi}|\partial_j\tilde{\psi}\rangle-\langle\tilde{\psi}|\partial_i\tilde{\psi}\rangle\langle\partial_j\tilde{\psi}|\tilde{\psi}\rangle\right)\dif R^i\dif R^j.
\end{align}
It can be verified that the new metric
\begin{align}\label{QGT0}
g_{ij}=\langle\partial_i\tilde{\psi}|\partial_j\tilde{\psi}\rangle-\langle\tilde{\psi}|\partial_i\tilde{\psi}\rangle\langle\partial_j\tilde{\psi}|\tilde{\psi}\rangle
\end{align}
is gauge invariant.
Therefore, $g_{ij}$ is a proper measure of the local distance between physically inequivalent states in the quantum phase space $P(\mathcal{H})$. The gauge-invariant metric is also referred to as the quantum geometric tensor (QGT) \cite{QGTCMP80,QGT10}.

\subsection{K$\ddot{\text{a}}$hler manifold and QGT}\label{IID}
To gain a deeper understanding of the geometric foundation of the QGT and its link to the K$\ddot{\text{a}}$hler geometry, we present an alternative and systematic construction.
We consider the state $|\psi\rangle\in P(\mathcal{H})$ depending on the real-valued parameters $\mathbf{R}=(R^1,R^2,\cdots,R^k)^T$ and $X(t)$ being the tangent vector of the curve $\mathbf{R}(t)$. This implies $|X(t)\rangle=\frac{\dif}{\dif t}|\psi(\mathbf{R}(t))\rangle=|\partial_i\psi\rangle\dot{R}^i$. By using $X$ in the gauge-invariant inner product summarized in Appendix~\ref{app:inner}, we have
\begin{align}\label{FSm1}
\langle \partial_i\psi|\partial_j\psi\rangle=\langle \partial_i\tilde{\psi}|\partial_j\tilde{\psi}\rangle-\langle \partial_i\tilde{\psi}|\tilde{\psi}\rangle\langle\tilde{\psi}|\partial_j\tilde{\psi}\rangle.
\end{align}
which is exactly the QGT of Eq.~(\ref{QGT0}).
This means that the metric induced in Eq.~(\ref{qd3}) is indeed a metric on $P(\mathcal{H})=CP^{N-1}$. Explicitly,
\begin{align}\label{dCPN}
\dif s^2(CP^{N-1})&=\big||\psi(\mathbf{R}+\dif \mathbf{R})\rangle-|\psi(\mathbf{R})\rangle\big|^2\notag\\&=g_{ij}\dif R^i\dif R^j\equiv\langle \partial_i\psi|\partial_j\psi\rangle\dif R^i\dif R^j
\end{align}
is the quantum distance between the equivalent-classes represented by $|\psi(\mathbf{R}+\dif \mathbf{R})\rangle$ and $|\psi(\mathbf{R})\rangle$. Note $g_{ij}$ is complex in general, and its real part can be obtained simply by symmetrizing $i$ and $j$ since $\dif R^i\dif R^j$ are symmetric about $i\leftrightarrow j$. Using the fact $\langle\partial_i\tilde{\psi}|\tilde{\psi}\rangle=-\langle\tilde{\psi}|\partial_i\tilde{\psi}\rangle$, we have
\begin{align}\label{h}
\gamma_{ij}\equiv\text{Re}g_{ij}=\text{Re}\langle \partial_i\tilde{\psi}|\partial_j\tilde{\psi}\rangle-\langle \partial_i\tilde{\psi}|\tilde{\psi}\rangle\langle\tilde{\psi}|\partial_j\tilde{\psi}\rangle.
\end{align}
This is the aforementioned Riemannian metric on $P(\mathcal{H})=CP^{N-1}$ that satisfies $\dif s^2(CP^{N-1})=\sum_{ij}\gamma_{ij}\dif R^i\dif R^j$.
In other words, the imaginary part of $g_{ij}$ makes no contribution to the quantum distance.

By anti-symmetrizing Eq.~(\ref{FSm1}), we get the (negative) imaginary part of $g_{ij}$: \begin{align}\label{O}\Omega_{ij}=-\text{Im}\langle \partial_i\tilde{\psi}|\partial_j\tilde{\psi}\rangle=\frac{\mi}{2}\left(\langle \partial_i\tilde{\psi}|\partial_j\tilde{\psi}\rangle-\langle \partial_j\tilde{\psi}|\partial_i\tilde{\psi}\rangle\right)\end{align} and the symplectic form
\begin{align}\label{Oeb}
\Omega_{|\psi\rangle}=\frac{\mi}{2}\langle \partial_i\tilde{\psi}|\partial_j\tilde{\psi}\rangle\dif R^i\wedge\dif R^j.
\end{align}
It can be verified that both $\gamma_{ij}$ and $\Omega_{ij}$ are gauge invariant.
Using $\langle\partial_i\partial_k\tilde{\psi}|\cdot\rangle\dif R^k\wedge\dif R^i\wedge\cdot=0=\langle\cdot|\partial_i\partial_k\tilde{\psi}\rangle\dif R^k\wedge\dif R^i\wedge\cdot$, it can be shown $\dif\Omega_{|\psi\rangle}=0$, i.e., the symplectic form is closed. Therefore, $P(\mathcal{H})=CP^{N-1}$ is indeed a K$\ddot{\text{a}}$hler manifold and $\Omega_{|\psi\rangle}$ is  called the K$\ddot{\text{a}}$hler form \cite{Nakahara}. More details of the K$\ddot{\text{a}}$hler manifold and its metric are given in Appendix~\ref{appKahler}.

Introducing the Berry connection $A=A_i\dif R^i$, whose $i$th component is $A_i=\langle \psi|\partial_i\psi\rangle$, and explicitly expressing $|\tilde{\psi}\rangle=\me^{\mi\theta}|\psi\rangle$, we find \begin{align}\label{tmp1}\langle \partial_i\tilde{\psi}|\tilde{\psi}\rangle\langle\tilde{\psi}|\partial_j\tilde{\psi}\rangle=\langle \partial_i\psi|\psi\rangle\langle\psi|\partial_j\psi\rangle\end{align} since the derivatives of $\theta$ cancel each other. Thus, the QGT can be written as
\begin{align}\label{FSm1b}
g_{ij}=\langle \partial_i\tilde{\psi}|\partial_j\tilde{\psi}\rangle+A_iA_j
\end{align}
according to Eq.~(\ref{FSm1}).
We emphasize that $A=\langle \psi|\dif\psi\rangle$ introduced here is the Berry connection on $P(\mathcal{H})$, while Eq.~(\ref{BA1}) defines the Berry connection $\omega$ on $S(\mathcal{H})$. The relation between them will be discussed later.
It is straightforward to verify from Eq.~(\ref{Oeb}) that the K$\ddot{\text{a}}$hler form is
\begin{align}
\Omega&=\frac{\mi}{2}\langle \dif\psi|\wedge|\dif\psi\rangle+\frac{\mi}{2}\dif\theta\wedge \dif \theta \notag\\
&+\frac{1}{2}\left(\dif \theta\wedge \langle\psi|\dif \psi\rangle-\langle\dif\psi|\psi\rangle\wedge \dif\theta\right)\notag\\
&=\frac{\mi}{2}\dif A.
\end{align}
Thus, the imaginary part is proportional to the gauge-invariant Berry curvature $F=\dif A$, so no modification is needed.

Eq.~(\ref{qd3}) thus represents the local distance between $|\psi(\mathbf{R}+\dif \mathbf{R})\rangle$ and $|\psi(\mathbf{R})\rangle$ in $CP^{N-1}$, as explained by Eq.~(\ref{dCPN}). Similarly, the distance of Eq.~(\ref{qd1}) may be viewed as the `raw' distance between $|\tilde{\psi}(\mathbf{R}+\dif \mathbf{R})\rangle$ and $|\tilde{\psi}(\mathbf{R})\rangle$ in $S(\mathcal{H})=S^{2N-1}$. However the states may be physically equivalent to each other. We denote the raw distance by
\begin{align}\dif s^2(S^{2N-1})=\langle \partial_i\tilde{\psi}|\partial_j\tilde{\psi}\rangle \dif R^i\dif R^j,
\end{align}
where $\langle \partial_i\tilde{\psi}|\partial_j\tilde{\psi}\rangle$ is the `raw' metric.
The Berry connection $A$ builds a bridge that connects $\dif s^2(S^{2N-1})$ and $\dif s^2(CP^{N-1})$. We notice from Eq.~(\ref{FSm1b}) that
\begin{align}\label{FSm1c}
\frac{\dif s^2(CP^{N-1})}{\dif t^2}&=\langle \partial_i\tilde{\psi}|\partial_j\tilde{\psi}\rangle \dot{R}^i\dot{R}^j+(\langle\tilde{\psi}|\partial_i\tilde{\psi}\rangle\dot{R}^i)^2 \notag\\&\le
\frac{\dif s^2(S^{2N-1})}{\dif t^2}.
\end{align}
The inequality is due to $\langle\tilde{\psi}|\partial_i\tilde{\psi}\rangle$ being a purely imaginary number, so $ (\langle\tilde{\psi}|\partial_i\tilde{\psi}\rangle\dot{R}^i)^2$ is negative.
Hence, $\dif s^2(S^{2N-1})$ is minimized if
\begin{align}\label{FSm1d}
\langle\tilde{\psi}|\partial_i\tilde{\psi}\rangle\dot{R}^i=\langle\tilde{\psi}(t)|\frac{\dif}{\dif t}|\tilde{\psi}(t)\rangle=\text{Im}\langle\tilde{\psi}(t)|\frac{\dif}{\dif t}|\tilde{\psi}(t)\rangle=0,
\end{align}
where the second identity comes from  $\langle\tilde{\psi}|\tilde{\psi}\rangle=1$, and `d' is the local derivative on $S(\mathcal{H})$. We previously show that $\frac{\dif}{\dif t}|\tilde{\psi}(t)\rangle$ is actually $|\tilde{X}(t)\rangle$. 
Hence, the condition (\ref{FSm1c}) simply implies that $\langle\tilde{\psi}(t)|\tilde{X}(t)\rangle=0$ or $|\tilde{X}\rangle=|\tilde{X}^\perp\rangle$, indicating that
\begin{align}
\langle\tilde{\psi}(t)|\tilde{\psi}(t+\dif t)\rangle&=\langle\tilde{\psi}(t)|\tilde{\psi}(t)\rangle+\langle\tilde{\psi}(t)\frac{\dif}{\dif t}|\tilde{\psi}(t)\rangle\dif t+\cdots\notag\\
&\approx 1>0.
\end{align}

In quantum information \cite{WatrousBook,Uhlmann86}, the above result means $|\tilde{\psi}(t+\dif t)\rangle$ is parallel to $|\tilde{\psi}(t)\rangle$ since they are `in phase' with each other. As a consequence, Eq.~(\ref{FSm1d}) is the parallel-transport condition, and $|\tilde{\psi}(t)\rangle$ is referred to as the horizontal lift of $|\psi(t)\rangle$. However, the parallelism between quantum states does not define an equivalence relation because it lacks transitivity \cite{OurPRB20b}. Therefore, even if $|\tilde{\psi}\rangle$ is parallel-transported, it may gradually start to be out of phase with the initial state, and the Berry phase is a measure of this discrepancy for a cyclic evolution. Substituting $|\tilde{\psi}(t)\rangle=\me^{\mi\theta(t)}|\psi(t)\rangle$ into Eq.~(\ref{FSm1d}) gives
 \begin{align}\label{BPen}
\mi\dot{\theta}+\langle\psi(t)|\frac{\dif}{\dif t}|\psi(t)\rangle=0.
\end{align}
If $|\psi(t)\rangle$ undergoes a cyclic process ($|\psi(0)\rangle=|\psi(\tau)\rangle$) that lasts for a duration $\tau$ ($0\le t\le \tau$), in the end the state obtains a gauge-invariant Berry phase
\begin{align}\label{BP}
\theta_B\equiv \theta(\tau)=\mi\int_0^\tau\dif t\langle\psi(t)|\frac{\dif}{\dif t}|\psi(t)\rangle.
\end{align}

Using the expression of the Berry connection $A=\langle \psi|\dif \psi\rangle$ and $|\tilde{\psi}\rangle=\me^{\mi\theta}|\psi\rangle$, we have
\begin{align}\label{FSm1c3}
\langle \tilde{\psi}|\partial_i\tilde{\psi}\rangle\dif R^i=\langle \tilde{\psi}|\dif \tilde{\psi}\rangle=\mi(\dif\theta-\mi A).
\end{align}
The first line of Eq.~(\ref{FSm1c}) then indicates   \begin{align}\label{FSm1c4}
\dif s^2(S^{2N-1})=\dif s^2(CP^{N-1})+(\dif\theta-\mi A)^2.
\end{align}
This gives a Pythagorean-like relation for the quantum distances, which is a more complete relation than  \begin{align}\label{FSm1c4d}
\dif s^2(CP^{N-1})=\inf [\dif s^2(S^{2N-1})].
\end{align}
We remark that a similar result for unnormalized quantum states was given in Ref.~\cite{PhysRevA.36.3479} obtained by using complex local coordinates of $\mathbb{C}^N$.
In the fiber-bundle description of the Berry phase \cite{PhysRevA.36.3479,OurDPUP}, $S^{2N-1}$ is the total space of the Hopf bundle over the base space $CP^{N-1}$. All possible phase factors at $|\psi\rangle\in CP^{N-1}$ form a fiber space at that point. Thus, Eq.~(\ref{FSm1c4}) shows the local distance on $S^{2N-1}$ has two contributions respectively from the base and the fiber spaces. During a physical process, if the quantum state undergoes parallel transport depicted in Eq.~(\ref{BPen}), no contribution from the U(1) phase factor adds to the total distance since the state is kept `in phase' instantaneously. In this situation, $\dif s^2(S^{2N-1})$ is minimized. This is indeed the definition of the Fubini-Study distance between quantum states \cite{Uhlmann95}:
\begin{align}\label{dFS}
\dif s^2_\text{FS}(|\psi(t+\dif t)\rangle,|\psi(t)\rangle)&:=\inf\big||\tilde{\psi}(t+\dif t)\rangle-|\tilde{\psi}(t)\rangle\big|^2 \notag \\
&=2-2\langle\tilde{\psi}(t+\dif t)|\tilde{\psi}(t)\rangle.
\end{align}
The infimum is obtained when $|\tilde{\psi}(t+\dif t)\rangle\px |\tilde{\psi}(t)\rangle$ or equivalently  $\langle\tilde{\psi}(t)|\tilde{\psi}(t+\dif t)\rangle=\langle\tilde{\psi}(t+\dif t)|\tilde{\psi}(t)\rangle>0$. This is why the metric (or the QGT) given by Eq.~(\ref{FSm1b}) is also called the Fubini-Study metric \cite{EGUCHI1980213}.

To summarize, the real part of the QGT, $\gamma_{ij}$, measures the local distance of a parameter-dependent state $|\psi(t)\rangle$ in the quantum phase space $P(\mathcal{H})$ while the imaginary part $\Omega_{ij}$ of the QGT characterizes the violation of parallelism between quantum states during a cyclic parallel-transport and the local curvature of the evolutional curve $|\psi(t)\rangle$. We emphasize that $P(\mathcal{H})=CP^{N-1}$ possesses those geometric properties because it is a K$\ddot{\text{a}}$hler manifold, and a more detailed discussion is outlined in Appendix \ref{appKahler}.

\section{QGT of mixed states}\label{SecIII}
Mixed states are inevitable for systems at finite temperatures or out of equilibrium. It is thus natural and necessary to extend the formalism of the QGT to cover mixed-state scenarios. Since the QGT is a measure of the local geometry of quantum states, it is important to analyze the properties of the space(s) formed by mixed states.

\subsection{Phase space of mixed states}
A mixed quantum state, also known as a statistical ensemble, is a collection of distinct pure states $\{|i\rangle\}$, each with an associated, non-negative probability $\lambda_i$. In quantum mechanics, this ensemble $\{(\lambda_i,|i\rangle)\}$ is usually represented by a density matrix, denoted by $\rho=\sum_i\lambda_i|i\rangle\langle i|$. Mathematically, $\rho$ is an $N\times N$ complex matrix satisfying three conditions: (1) Hermiticity: $\rho=\rho^\dag$, (2) non-negativity: $\rho\ge 0$ (all the eigenvalues of $\rho$ are non-negative), and (3) normalization: $\text{Tr}\rho=1$.
The mixed state is not in one-to-one correspondence with the density matrix, as stated by Schr$\ddot{\text{o}}$dinger's mixture theorem \cite{Bengtsson_book}. Interesting examples can be found in Ref.~\cite{MQM}.
In general, there may be infinitely many different ensembles represented by the same density matrix \cite{Kirkpatrick06}, but they cannot be distinguished by any physical measurement \cite{Ochs81}. Therefore, it is sufficient to use density matrices to represent physically distinguishable mixed states.

We denote by $\mathcal{P}$ the set, or `space', of all density matrices. Unlike the pure-state situation, $\mathcal{P}$ is neither a linear space nor a manifold but only a convex subset of $End(\mathcal{H})$, the set of endomorphisms (or operators) of $\mathcal{H}$~\cite{Bengtsson_book}.
Since $\mathcal{P}$ is not even a linear space, its dimension can not be defined in the usual way. Here we refer to it as the number $N^2-1$ of real parameters necessary to completely specify an arbitrary density matrix \cite{Bengtsson_book}. Some details can be found in Appendix \ref{app0}. Mathematically, $ \mathcal{P}$ can be thought of as a convex rigid body in $\mathds{R}^{N^2-1}$. Thus, we will restrict our discussion to $N>1$ since if $N=1$, $\mathds{R}^{0}$ is a single point, so is $\mathcal{P}$.

Although $\mathcal{P}$ is not a manifold, it is the union of a series of manifolds: $\mathcal{P}=\bigcup_{k=1}^N\mathcal{D}^N_k$.
Here $\mathcal{D}^N_k$ is the space of normalized $N\times N$ density matrices of fixed rank $k$ ($1\le k\le N$), which is a manifold equipped with a Riemannian metric \cite{REP95}. However, the metrics of $\mathcal{D}^N_k$ cannot be glued together to construct a metric of $\mathcal{P}$ due to the conical singularities in the neighborhood of the boundary of $ \mathcal{D}^N_N$ if $N>2$ \cite{REP95}. This is understandable since $\mathcal{P}$ is a positive cone in $End(\mathcal{H})$ \cite{Bengtsson_book}.
The density matrix of a pure state is a projective operator with $\rho^2=\rho$ and rank 1.
Thus, the set of pure-state density matrices is equivalent to $\mathcal{D}^N_1$, i.e., $D^N_1\simeq\{\rho\in \mathcal{P}|\rho^2=\rho\}$.
Geometrically, $D^N_1$ is equivalent to $CP^{N-1}$, the phase space of pure states, and they both have dimension $2(N-1)$.
For pure states, we introduced the manifold of all normalized states $S(\mathcal{H})=S^{2N-1}$ after the phase space $CP^{N-1}$ is defined, thereby obtaining the fibration $S^{2N-1}/\text{U}(1)=CP^{N-1}$ by collapsing all the U(1) phase factors. Here we ask whether a similar fibration also exists for mixed states with $\mathcal{D}^N_N$ as the phase space. If so, what plays the similar role of $S^{2N-1}$?

Those questions can be resolved by following Uhlmann's approach via purification \cite{Uhlmann86}.
The Uhlmann bundle was based on full-rank density matrices.
Therefore, we will focus exclusively on $\mathcal{D}^N_N$, the submanifold of all full-rank density matrices, which include all mixed states in thermal equilibrium. To investigate the local geometry of the Uhlmann bundle, we recognize $\mathcal{D}^N_N$ as the phase space of mixed states of interest.
Unlike the phase space of pure states that forms a K$\ddot{\text{a}}$hler manifold with many geometric properties, the mixed-state phase space $\mathcal{D}^N_N$ lacks some characteristics but still exhibits interesting local geometry, which will be shown shortly. The dimension of $\mathcal{D}^N_N$ is $N^2-1$ while $\dim \mathcal{D}^N_k=N^2-(N-k)^2-1$. A proof is outlined at the end of Appendix \ref{app0}.

An operator $W\in End(\mathcal{H})$ is called a purification or amplitude of $\rho\in \mathcal{D}^N_N$ iff $\rho=WW^\dag$, which implicitly requires that $W$ also has full rank.
Conversely, a full-rank matrix $W$ has a unique polar decomposition $W=\sqrt{\rho}U$, where $U\in$U$(N)$. The uniqueness of the decomposition of purification is another reason that we recognize $\mathcal{D}^N_N$ as the phase space of mixes states.
When compared with the pure-state case, $W$ and $U$ are the counterparts of the wavefunction and phase factor, respectively. Moreover, $\text{Tr}\rho=1$ leads to $\text{Tr}(W^\dag W)=1$. Hence, we introduce the Hilbert-Schmidt product of objects in $End(\mathcal{H})$:
\begin{align}
\langle A,B\rangle_\text{HS}=\text{Tr}(A^\dag B).
\end{align}
This leads us to further define the Hilbert-Schmidt norm: $||W||_\text{HS}=\langle W,W\rangle_\text{HS}$. With these notations, we introduce $\mathcal{S}_N\subset End(\mathcal{H})$ as the unit sphere with respect to the Hilbert-Schmidt norm, which is a manifold consisting of all full-rank (rank $N$) density matrices with unit norm.  $\mathcal{S}_N$ plays a similar role as its pure-state counterpart $S(\mathcal{H})=S^{2N-1}$. In this way, a fibration $\mathcal{S}_N/\text{U}(N)=\mathcal{D}^N_N$ for mixed states is established via the projection
\begin{align}\label{empi}
 \pi:\mathcal{S}_N \rightarrow \mathcal{D}^N_N,\quad  \pi(W)=WW^\dag=\rho.
\end{align}
Furthermore, $\rho$ is invariant under the gauge transformation $W\rightarrow W'=W\mathcal{U}$, where $\mathcal{U}\in$ U$(N)$.

While $W$ is a state-matrix, its counterpart $|\tilde{\psi}\rangle$ is a state-vector. The difference brings some inconvenience when making a direct comparison. The purified state associated with $W$ is introduced as follows.
If $\rho$ is diagonalized as  $\rho=\sum_i\lambda_i|i\rangle\langle i|$, $W=\sum_i\sqrt{\lambda_i}|i\rangle\langle i|U$ with the corresponding purified state defined as $|W\rangle=\sum_i\sqrt{\lambda_i}|i\rangle\otimes U^T|i\rangle$ by taking the transpose of `$\langle i|U$' of $W$. Thus, $|W\rangle \in \mathcal{H}\otimes\mathcal{H}^*$.
Finally, an inner product between two purified states is introduced as $\langle W_1|W_2\rangle=\langle W_1,W_2\rangle_\text{HS}=\text{Tr}(W_1^\dag W_2)$ \cite{HubnerPLA93}, which preserves the Hilbert-Schmidt product.

\subsection{The Hilbert-Schmidt metric}
To generalize the concept of QGT to mixed quantum states, it is necessary to give a suitable definition for the quantum distance between two physically distinct mixed states. 
We have introduced the Hilbert-Schmidt product on $End(\mathcal{H})$ previously, which leads to the Hilbert-Schmidt distance
\begin{align}
\dif^2_\text{HS}(A,B)=\frac{1}{2}\text{Tr}\left[(A-B)(A^\dag- B^\dag)\right].
\end{align}
Since $\rho=\rho^\dag$, the Hilbert-Schmidt distance between two density matrices $\rho$ and $\rho'$ is given by
\begin{align}\label{dHSrho}
\dif^2_\text{HS}(\rho,\rho')=\frac{1}{2}\text{Tr}(\rho-\rho')^2.
\end{align}
Let $\rho'\rightarrow \rho+\dif\rho $, the infinitesimal Hilbert-Schmidt distance becomes $\dif^2_\text{HS}(\rho,\rho+\dif \rho)=\frac{1}{2}\text{Tr}(\dif\rho)^2$.
We choose $\mathbf{R}=(R^1,R^2,\cdots,R^k)^T$ as before, then $\dif\rho=\partial_\mu\rho\dif R^\mu$ and the corresponding Hilbert-Schmidt metric is \begin{align}g^\text{HS}_{\mu\nu}=\frac{1}{2}\text{Tr}(\partial_\mu\rho\partial_\nu\rho).\end{align}
However, $g^\text{HS}_{\mu\nu}$ may not be an appropriate choice to capture the local geometry of $\mathcal{D}^N_N$ since it is equivalent to a flat metric on $End(\mathcal{H})$, which will be shown below.

As explained previously, $N^2-1$ parameters are needed to specify a full-rank density matrix $\rho$. Therefore, with the help of the Bloch vector $\mathbf{a}(\mathbf{R})=(a^1(\mathbf{R}),a^2(\mathbf{R}),\cdots, a^{N^2-1}(\mathbf{R}))^T$, any density matrix can be decomposed as \cite{Bengtsson_book}
\begin{align}\label{dmd1b}
\rho=\frac{1}{N}1_N+\sum_{i=1}^{N^2-1}a^i(\mathbf{R})T_i.
\end{align}
Here $1_N$ is the $N\times N$ identity matrix, and $T_i$s are the generators of $\text{SU}(N)$ that satisfy
\begin{align}\label{4.20}T_iT_j=\frac{2}{N}\delta_{ij}+\sum_kd_{ijk}T_k+\mi \sum_kf_{ijk}T_k.\end{align}
Here $d_{ijk}$ is a totally symmetric tensor that can only be defined if $N>2$, and $f_{ijk}$ is a totally anti-symmetric tensor. Using these properties and $\text{Tr}T_i=0$, it can be shown that $\dif_\text{HS}^2(\rho,\rho')=\dif_\text{HS}^2(\sum_ia^i\sigma_i,\sum_ia'^i\sigma_i')=\sum_i(a^i-a'^i)^2$. Hence,
\begin{align}\label{dHSrho2}
\dif_\text{HS}^2&(\rho,\rho+\dif\rho)=\dif\mathbf{a}\cdot\dif\mathbf{a}=\sum_{ij}\delta_{ij}\dif a^i\dif a^j,
\end{align}
which exhibits the Euclidean-like distance. Accordingly, the components $a_i$ also serves as the Cartesian coordinates in Bloch space, and the induced metric $\delta_{ij}$ from Eq.~(\ref{dHSrho2}) is flat. Thus, it does not reveal interesting properties of  $\mathcal{D}^N_N$ since it cannot accurately capture the local geometry. In terms of $\mathbf{R}$, the Hilbert-Schmidt metric is
\begin{align}g^\text{HS}_{\mu\nu}=\frac{\partial \mathbf{a}}{\partial R^\mu}\cdot\frac{\partial \mathbf{a}}{\partial R^\nu}=\frac{\partial a^i}{\partial R^\mu}\frac{\partial a^j}{\partial R^\nu}\delta_{ij},
\end{align}
which is simply the coordinate-transformed expression of the flat metric.

\subsection{Purification and Bures metric}\label{SecIIIC}
To construct a suitable Riemannian metric on $\mathcal{D}^N_N$, we need to find a way to isometrically embed $\mathcal{D}^N_N$ into $End(\mathcal{H})$, or more accurately, into $\mathcal{S}_N$ since $\text{Tr}\rho=\langle W,W\rangle_\text{HS}=1$. This is achievable through the approaches by Uhlmann \cite{Uhlmann86,Uhlmann89} and Dittmann~\cite{REP95}. However, their original formulations were quite mathematical. Here we combine their ideas with a more pedagogical description to elucidate the geometrical meaning and physical implications.
The projection (\ref{empi}) provides an embedding $\rho=\pi(W)$, which may induce a metric from the Hilbert-Schmidt metric via its push-forward map $\pi_*$~\cite{REP95}.
The metric $g$ on $\mathcal{D}^N_N$ is a type (0,2) tensor field \cite{Nakahara} defined by $g(\cdot,\cdot):T_\rho\mathcal{D}^N_N\times T_\rho\mathcal{D}^N_N\rightarrow \mathbb{C}$, similar to Eq.~(\ref{Hip}). Explicitly, if $Y_\rho$ is a tangent vector at the point $\rho$, i.e., $Y_\rho\in T_\rho\mathcal{D}^N_N$, then
\begin{align}\label{gyy}
 g(Y_\rho,Y_\rho):=\inf\{\langle V,V\rangle_\text{HS}|\pi_*(V)=Y_\rho,V\in T\mathcal{S}_N\}.
\end{align}
Since the Hilbert-Schmidt metric is invariant under the gauge transformation $V\rightarrow V\mathcal{U}$ with $\mathcal{U} \in \text{U}(N)$, it is sufficient to take $V$ tangent to $\mathcal{S}_N$ at $W(0)=\sqrt{\rho}$.
We consider a curve parameterized by $t$ starting from $\rho$: $\rho(t)=\rho(0)+t Y_\rho$, where $t$ is small and $\rho(0)=\rho$. Based on the above assumptions, the purification of the density matrix is $W(t)=W(0)+tV$ with $W(0)=\sqrt{\rho}$. Using $\rho(t)=W(t)W^\dag(t)$, we obtain $\sqrt{\rho}V^\dag+V\sqrt{\rho}=Y_\rho$. Moreover, since $V=\lim_{t\rightarrow 0}\frac{W(t)-W(0)}{t}$, Eq. (\ref{gyy}) becomes
\begin{widetext}
\begin{align}\label{gyy2}
 g(Y_\rho,Y_\rho):=\inf\{\lim_{t\rightarrow 0}\frac{1}{t^2}\langle W(t)-W(0),W(t)-W(0)\rangle_\text{HS}|\sqrt{\rho}V^\dag+V\sqrt{\rho}=Y_\rho\}.
\end{align}
\end{widetext}
This is equivalent to the infimum of the Hilbert-Schmidt distance between $W(t)$ and $W(0)=\sqrt{\rho}$, which is
the Bures distance between $\rho(t)$ and $\rho(0)=\rho$ \cite{Uhlmann86}. Explicitly,
\begin{align}\label{gyy3}
&\dif^2_\text{B}(\rho(t),\rho):=\inf\text{Tr}\left(W(t)-W(0)\right)\left(W(t)-W(0)\right)^\dag\notag\\
=&\inf\left[2-\text{Tr}(W^\dag(0)W(t)+W^\dag(t)W(0))\right],
\end{align}
where the infimum is obtained with respect to all possible $W(t)$ such that $\rho(t)=W(t)W^\dag(t)$.
From this definition, the Bures metric is indeed the desired metric when $\mathcal{D}^N_N$ is isometrically embedded. It should be noted that the infimum is taken among all possible purifications, thus it is independent of the gauge choice of $W$. In other words, it is by definition invariant under the gauge transformation $W\rightarrow W'=W\mathcal{U}$, and the same argument applies to the Bures metric.

The right-hand-side of Eq.~(\ref{gyy3}) takes the infimum
\begin{align}\label{db3}
\dif^2_\textrm{B}(\rho(t),\rho)=\dif^2_\textrm{B}(\rho,\rho(t))=2-2\textrm{Tr}\sqrt{\sqrt{\rho}\rho(t)\sqrt{\rho}}
\end{align}
 iff (see Appendix \ref{appa0})
 \begin{align}\label{pcm}
W^\dag(t)W(0)=W^\dag(0)W(t)>0,
\end{align}
where $\textrm{Tr}\sqrt{\sqrt{\rho}\rho(t)\sqrt{\rho}}$ is called the Uhlmann fidelity \cite{HubnerPLA93,Uhlmann95}, and Eq.~(\ref{pcm}) is called the Uhlmann parallel condition between $W(t)$ and $W(0)$~\cite{Uhlmann86}. This condition implies that the tangent vector $V$ is also parallel to $W(0)$. Furthermore, how $V$ leads to the infimum of (\ref{gyy2}) is shown in Appendix \ref{appa0}.  Analogous to the situation of pure states, $V\px W(0)=\sqrt{\rho}$ means $V$ belongs to the horizontal subspace of $T_{W(0)}\mathcal{S}_N$. We will revisit this topic later.
The differential form of Eq.~(\ref{pcm}) is
\begin{align}\label{pcm2}
\dot{W}^\dag(t)W(t)\big|_{t\rightarrow 0}=W^\dag(t)\dot{W}(t)\big|_{t\rightarrow 0}.
\end{align}
Taking trace of both sides, we get a necessary condition in terms of purified states
\begin{align}\label{pcm3}
\text{Im}\langle W(t)|\frac{\dif}{\dif t}|W(t)\rangle=0,
\end{align}
which can be viewed as a generalization of Eq.~(\ref{FSm1d}) for pure states. However, this is only a weaker and necessary condition of Eq.~(\ref{pcm2}).

The Bures distance in terms of the generic local coordinates $\mathbf{R}=(R^1,R^2,\cdots,R^k)^T$ is
 \begin{align}\label{A1}
\dif^2_\text{B}(\rho,\rho+\mathrm{d}\rho)=g^\text{B}_{\mu\nu}(\rho(\mathbf{R}))\mathrm{d}R^\mu\mathrm{d}R^\nu,
\end{align}
where $g^\text{B}_{\mu\nu}$ is the Bures metric.
Using Eq.~(\ref{db3}), it can be found that \cite{HubnerPLA92}
 \begin{align}\label{A11}
\dif^2_\text{B}(\rho,\rho+\mathrm{d}\rho)=&\frac{1}{2}\sum_{ij}\frac{|\langle i|\mathrm{d}\rho|j\rangle|^2}{\lambda_i+\lambda_j}\notag\\=&\frac{1}{2}\sum_{ij}\frac{\langle i|\partial_\mu\rho|j\rangle\langle j|\partial_\nu\rho|i\rangle}{\lambda_i+\lambda_j}\dif R^\mu\dif R^\nu.
\end{align}
Some details can be found in Appendix \ref{appa}.
Thus, the Bures metric is
 \begin{align}\label{Bm}
 g^\text{B}_{\mu\nu}=\frac{1}{2}\sum_{ij}\frac{\langle i|\partial_\mu\rho|j\rangle\langle j|\partial_\nu\rho|i\rangle}{\lambda_i+\lambda_j}.
\end{align}
Since the density matrix is invariant under the gauge transformation $W\rightarrow W'=W\mathcal{U}$, the Bures metric remains unchanged as well. Therefore, it is naturally a good candidate as the QGT for mixed states. Since Eq.~(\ref{A11}) represents the local distance on $\mathcal{D}^N_N$, we will interchangeably use $\dif^2_\text{B}(\mathcal{D}^N_N)$ to denote it hereafter.

\subsection{The Uhlmann metric}
So far, our discussions have revealed some similarities between the Fubini-Study metric for pure states and the Bures metric for mixed states.
The definition (\ref{gyy3}) of the Bures distance can be written as
 \begin{align}\label{db4}
\dif^2_\text{B}(\rho(t),\rho)=\inf ||W(t)-W(0)||_\text{HS}.
\end{align}
In terms of purified states, it is also expressed as
 \begin{align}\label{db5}
\dif^2_\text{B}(\rho(t),\rho)=\inf \big||W(t)\rangle-|W(0)\rangle\big|^2.
\end{align}
As a comparison, the results in Sec.~\ref{IID} show that
 \begin{align}\label{dFS3}
&\dif s^2_\text{FS}(|\psi(t)\rangle,|\psi(0)\rangle)=\inf \big||\tilde{\psi}(t)\rangle-|\tilde{\psi}(0)\rangle\big|^2.
\end{align}
The infima of the above expressions are attained when the parallel-transport conditions for $|\tilde{\psi}(t)\rangle$ and $W(t)$ are fulfilled, respectively. However, the infimum of Eq.~(\ref{db5}) is obtained with respect to all possible U($N$) phase factors $U(t)$ in $W(t)=\sqrt{\rho(t)}U(t)$ whereas that of Eq.~(\ref{dFS3}) is obtained with respect to all possible U(1) phase factors $\me^{\mi\theta(t)}$ in $|\tilde{\psi}(t)\rangle=\me^{\mi\theta(t)}|\psi(t)\rangle$. Thus, despite the superficial similarity, they are not equivalent. Moreover, Eqs.~(\ref{db5}) and (\ref{dFS3}) further lead to the Fubini-Study and Bures metrics, which are respectively invariant under the U(1) and U($N$) gauge transformations. The gauge invariance makes them suitable choices for the QGTs in their respective situations.

For pure states, however, the Fubini-Study metric can be derived either through a gauge-invariant modification to the raw metric $\langle \partial_i\tilde{\psi}|\partial_i\tilde{\psi}\rangle$ or by directly constructing a gauge-invariant metric on $CP^{N-1}$. Subsequently, it was found that the requirement of gauge-invariance saturates the inequality $\dif s^2(CP^{N-1})\le \dif s^2(S^{2N-1})$. In addition, the local geometry associated with the fibration $S^{2N-1}/\text{U}(1)=CP^{N-1}$ is manifested by the relation (\ref{FSm1c4}). For mixed states, the Bures metric is obtained by taking the infimum of the Hilbert-Schmidt distance between two purifications, which is guaranteed to be gauge-invariant by definition. Despite the progress made so far, there still remain some intriguing puzzles that need to be addressed. For example, is there a similar relation associated with the fibration $\mathcal{S}_N/\text{U}(N)=\mathcal{D}^N_N$ just like the case of pure states? Additionally, the parallel-transport condition (\ref{pcm}) minimizes the right-hand-side of Eq.~(\ref{db4}). Is it equivalent to the criterion for the metric to be gauge-invariant? Can there be other gauge-invariant metrics besides the Bures metric?

To answer these questions, we follow the same procedure as that of the pure-state case to modify the `raw metric' on $\mathcal{S}_N$.
Let $W(t)$ or $|W(t)\rangle$ depend on $t$ via the parameter curve $\mathbf{R}(t)$. From Eq.~(\ref{db4}), the local distance on $\mathcal{S}_N$ can be equivalently written as
\begin{align}\label{ds2}
&\dif s^2(\mathcal{S}_N)=\big|| W(\mathbf{R}+\dif \mathbf{R})\rangle-|W(\mathbf{R})\rangle\big|^2\notag\\=&\langle \partial_\mu W|\partial_\nu W\rangle\dif R^\mu\dif R^\nu=\text{Tr}(\partial_\mu W^\dag\partial_\nu W)\dif R^\mu\dif R^\nu,
\end{align}
where $\langle \partial_\mu W|\partial_\nu W\rangle$ is recognized as the desired raw metric. It is not invariant under the U($N$) gauge transformation $W'=W\mathcal{U}$. By symmetrizing the indices $\mu$ and $\nu$, the local distance takes the form $\dif s^2(\mathcal{S}_N)=\gamma_{\mu\nu}\dif R^\mu\dif R^\nu$, where $\gamma_{\mu\nu}=\frac{1}{2}\left(\langle \partial_\mu W|\partial_\nu W\rangle+\langle \partial_\nu W|\partial_\mu W\rangle\right)$ is the real part of $\langle \partial_\mu W|\partial_\nu W\rangle$. Similarly, the imaginary part $\sigma_{\mu\nu}=\frac{1}{2\mi}\left(\langle \partial_\mu W|\partial_\nu W\rangle-\langle \partial_\nu W|\partial_\mu W\rangle\right)$ gives no contribution to $\dif s^2(\mathcal{S}_N)$.
It can be verified that both $\gamma_{\mu\nu}$ and $\sigma_{\mu\nu}$ are not gauge invariant.
Based on the previous discussions, it can be found that $\gamma_{\mu\nu}$ reduces to the Bures metric if $W$ satisfies the condition (\ref{pcm2}). Here we ask if a more generic gauge-invariant metric can be constructed by imposing a suitable modification to $\gamma_{\mu\nu}$. Moreover, the imaginary part in the case of pure states is proportional to the gauge-invariant Berry curvature and needs no modification. In the case of mixed states, does a similar assertion hold for $\sigma_{\mu\nu}$ if it is appropriately modified?

We utilize Uhlmann's approach on geometric phase of mixed states \cite{Uhlmann86,Uhlmann89} to address the above questions.
Here we briefly outline Uhlmann's formalism as more recent overviews can be found in Refs. \cite{GPbook,TDMPRB15,OurDPUP,10.21468/SciPostPhysCore.6.1.024}. The theory of the Uhlmann phase is built on a U($N$) principle bundle, in which $\mathcal{S}_N$ is the total space, $\mathcal{D}^N_N$ is the base manifold, and the embedding $\pi$ defined by Eq.~(\ref{empi}) projects $\mathcal{S}_N$ to $\mathcal{D}^N_N$. As a generalization to the Berry connection, the Uhlmann connection $A_\text{U}$ is defined on $\mathcal{D}^N_N$ and is given by
 \begin{align}\label{Au0}
A_\text{U}=-\sum_{ij}|i\rangle\frac{\langle i|[\dif\sqrt{\rho},\sqrt{\rho}]|j\rangle}{\lambda_i+\lambda_j}\langle j|,
\end{align}
where `d' is the derivative on $\mathcal{D}^N_N$. It can be derived from the relation \cite{OurDPUP,10.21468/SciPostPhysCore.6.1.024}
\begin{align}\label{AU6'}
\rho \pi^*A_\text{U}+\pi^*A_\text{U}\rho=-[\dif_{\mathcal{S}_N}\sqrt{\rho},\sqrt{\rho}].
\end{align}
Here $\dif_{\mathcal{S}_N}$ is the derivative on the total space $\mathcal{S}_N$.
Similarly, the Ehresmann connection $\omega$ is defined on $\mathcal{S}_N$, which is related to $A_\text{U}$ via the pull-back $\pi^*$. Explicitly,
\begin{align}\label{AU4}
\omega=U^\dagger \pi^*A_\text{U}U+U^\dagger\dif_{\mathcal{S}_N} U.
\end{align}
Here $U$ is the phase factor of $W=\sqrt{\rho}U$. If a mixed state is transported along a curve $W(t)$ ($0\le t\le \tau$) with its tangent vector given by $\tilde{X}$, the parallel-transport condition (\ref{pcm2}) can be written as $\omega(\tilde{X})=0$ \cite{OurDPUP,10.21468/SciPostPhysCore.6.1.024}. Using Eq. (\ref{AU4}),  this is equivalent to
 \begin{align}\label{AU4b}
A_\text{U}(X)+\dot{U}U^\dag&=0,
\end{align}
or $A_{\text{U}\mu}+\partial_\mu UU^\dag=0$ in the component form,
where $X=\pi_*\tilde{X}=\frac{\dif}{\dif t}$.

We refer to any cyclic process during which a mixed state follow the Uhlmann parallel-transport condition as a Uhlmann process described by either Eq.~(\ref{pcm2}) or (\ref{AU4b}).
By solving Eq. (\ref{AU4b}), the initial and final phase factors of a Uhlmann process are off by an Uhlmann holonomy
\begin{align}\label{AU4c}
U(\tau)=\mathcal{P}\me^{-\oint A_\text{U}}U(0),
\end{align}
where $\mathcal{P}$ is the path ordering operator.
Similar to the Berry phase factor, the Uhlmann holonomy is a measure of the loss of parallelity during a cyclic parallel-transport.

It can be verified that under the gauge transformation $W'=W\mathcal{U}$, $\gamma_{\mu\nu}$, $\sigma_{\mu\nu}$, and $\omega$ change accordingly. A candidate of a gauge-invariant modification to $\gamma_{\mu\nu}$ is
\begin{align}\label{gmU}
g^\text{U}_{\mu\nu}:=\gamma_{\mu\nu}+\frac{1}{2}\text{Tr}\left(W^\dag  W\omega_\mu\omega_\nu+\omega_\nu\omega_\mu W^\dag W\right),
\end{align}
which is referred to as the Uhlmann metric.
Moreover, a gauge-invariant modification of the imaginary part leads to
\begin{align}\label{Uf}
\sigma^\text{U}_{\mu\nu}=\sigma_{\mu\nu}+\frac{\mi}{2}\text{Tr}\left[\partial_\mu(W^\dag W\omega_\nu)-\partial_\nu(W^\dag W\omega_\mu)\right].
\end{align}
One can verify that both $g^\text{U}_{\mu\nu}$ and $\sigma^\text{U}_{\mu\nu}$ are invariant under the gauge transformation $W'=W\mathcal{U}$. The proof is outlined in Appendix \ref{appc}. Their relations to the pure-state counterparts will be studied later.

\subsection{Uhlmann distance}
Since $g^\text{U}_{\mu\nu}$ is manifestly gauge-invariant, it is a candidate of a QGT of mixed states. However, what is the difference between the Uhlmann metric and the Bures metric? To facilitate a fair comparison, we restrict the components of $g^\text{U}_{\mu\nu}$ on $\mathcal{D}^N_N$. In other words, the previously chosen parameters $\mathbf{R}=(R^1,R^2,\cdots, R^k)^T$ are the local coordinates of $\mathcal{D}^N_N$. Then, $\omega_\mu=U^\dag A_{\text{U}\mu}U+U^\dag \partial_\mu U$. Using $W^\dag W=U^\dag \rho U$, the Uhlmann metric is given by
 \begin{align}\label{gm3}
g^\text{U}_{\mu\nu}=&\gamma_{\mu\nu}+\frac{1}{2}\text{Tr}\left[\rho(A_{\text{U}\mu}+\partial_\mu UU^\dag)(A_{\text{U}\nu}+\partial_\nu UU^\dag)\right]\notag\\+&\frac{1}{2}\text{Tr}\left[\rho(A_{\text{U}\nu}+\partial_\nu UU^\dag)(A_{\text{U}\mu}+\partial_\mu UU^\dag)\right].
\end{align}
With this expression, we can define the Uhlmann distance as
 \begin{align}\label{Ud}
\dif s^2_\text{U}=g^\text{U}_{\mu\nu}\dif R^\mu\dif R^\nu,
\end{align}

If we plug Eq.~(\ref{gmU}) into Eq.~(\ref{Ud}), an interesting result appears:
 \begin{align}\label{Ud2}
\dif s^2_\text{U}&=\langle \dif W|\dif W\rangle+\text{Tr}\left[\rho (A_\text{U}+\dif UU^\dag)^2\right]\notag\\
\Rightarrow \dif s^2(\mathcal{S}_N)&=\dif s^2_\text{U}+\text{Tr}\left[\rho (\mi A_\text{U}+\mi\dif UU^\dag)^2\right].
\end{align}
This is a Pythagorean-like equation for the distances of mixed states. In contrast to Eq.~\eqref{FSm1c4}, the mixed-state equation has not been discussed in the literature.
Here we emphasize that both $A_\text{U}$ and $\dif UU^\dag$ are anti-Hermitian, hence a factor $i$ from the rearrangement ensures that $(\mi A_\text{U}+\mi\dif UU^\dag)^2$ is positive-definite. Since the product of two positively-definite matrices is also a positively-definite matrix, the second term on the right-hand-side of the second line of Eq.~(\ref{Ud2}) is positive. To compare with the case of pure states, we rewrite Eq.~(\ref{FSm1c4}) by using $g=\me^{\mi\theta}$ or $\dif\theta=-\mi\dif g g^\dag$ to obtain
\begin{align}\label{FSm1c5}
\dif s^2(S^{2N-1})=\dif s^2(CP^{N-1})+(\mi A+\mi\dif g g^\dag)^2.
\end{align}
Interestingly, $\dif s^2_\text{U}$ plays a similar role as $\dif s^2(CP^{N-1})$, whose counterpart in mixed states is  the Bures distance. Moreover, Eq. (\ref{Ud2}) implies that when the parallel-transport condition ($A_{\text{U}\mu}+\partial_\mu UU^\dag=0$) is satisfied, $\dif s^2(\mathcal{S}_N)$ reduces to $\dif s^2_\text{U}$. However, the discussions in the previous subsection suggest that $\dif s^2(\mathcal{S}_N)$ reduces to the Bures distance under the same condition. Therefore, a reasonable guess is that the Uhlmann distance is equal to the Bures distance. Here we prove an equivalent statement: The Uhlmann metric on $\mathcal{D}^N_N$ reproduces the Bures metric, so the metric is unique in this sense.
We also remark that $\dif \theta$ is the differential angle or differential fiber $\mi\dif g g^\dag$ in Eq.~\eqref{FSm1c5} that the pure state obtains through parallel transport. Similarly, $A_\text{U}$ is the Uhlmann connection, and $\mi\dif UU^\dag$ is the change of the generator of the U($N$) phase factor. In pure states, $\dif\theta$ is the change of the generator of the U(1) phase factor. No loop integral of a cyclic process is involved, hence no geometric phase emerges, and only the local changes appear in Eqs.~\eqref{FSm1c4} and \eqref{FSm1c5}.

As summarized in Appendix~\ref{appa}, the Bures metric can be written as
\begin{align}\label{Bmb}
 g^\text{B}_{\mu\nu}=\frac{1}{2}\sum_{ij}\frac{(\sqrt{\lambda_i}+\sqrt{\lambda_j})^2}{\lambda_i+\lambda_j}\langle i|\partial_\mu\sqrt{\rho}|j\rangle\langle j|\partial_\nu\sqrt{\rho}|i\rangle.
\end{align}
Similarly, by expanding the commutator, the Uhlmann connection is expressed as
\begin{align}\label{Umr1}
A_{\text{U}\mu}=\sum_{ij}\frac{\sqrt{\lambda_i}-\sqrt{\lambda_j}}{\lambda_i+\lambda_j}\langle i|\partial_\mu\sqrt{\rho}|j\rangle|i\rangle\langle j|.
\end{align}
We show in Appendix~\ref{appd} that the Uhlmann distance is indeed equal to the Bures distance. Eq.~(\ref{Ud2}) thus takes the form
 \begin{align}\label{Ud3}
\dif s^2(\mathcal{S}_N)=\dif s^2_\text{B}(\mathcal{D}^N_N)+\text{Tr}\left[\rho (\mi A_\text{U}+\mi\dif UU^\dag)^2\right],
\end{align}
which is a manifestation of the fibration $\mathcal{S}_N/\text{U}(N)=\mathcal{D}^N_N$. This expression for mixed states serves as the counterpart of the relation (\ref{FSm1c5}) for pure states. A subtle difference is that $g$ is a U(1) phase factor, whereas $U$ is a U$(N)$ phase factor. The local distance on the total space $\mathcal{S}_N$ also has two contributions, respectively from the base manifold $\mathcal{D}^N_N$ and the fiber space U$(N)$. Similarly, the parallel transport of mixed states minimizes $\dif s^2(\mathcal{S}_N)$ since there is no `vertical' contribution from the fiber. The Bures metric is therefore the unique real part of the gauge-invariant QGT for mixed states within this framework. Moreover, Eq. (\ref{Ud3}) also indicates a crucial but previously unnoticed property of the geometry of mixed quantum states. We caution that  Eq. (\ref{Ud3}) is not a trivial generalization
because the Uhlmann connection does not become the Berry connection as $T\rightarrow 0$. The settlement of the uniqueness of the metric from the Uhlmann bundle not only provides a unified description of its local geometry but also dissuades people from futile trials of constructing alternative gauge-invariant metrics on the bundle.

\subsection{Sj$\ddot{\text{o}}$qvist distance}
Recently, another type of quantum distance between mixed states was introduced by E. Sj$\ddot{\text{o}}$qvist in Ref. \cite{PhysRevResearch.2.013344}, which will be referred to as the Sj$\ddot{\text{o}}$qvist distance for convenience. Its relation to the Bures distance can also be deduced via purification. Appendix~\ref{app:Sdistance} briefly reviews the construction of the Sj$\ddot{\text{o}}$qvist distance, which is given by
 \begin{align}\label{Sdis3}
\dif^2_\text{S}(\rho(t),\rho)=\inf_{\theta_n(t)}\big||W(t)\rangle-|W(0)\rangle\big|^2.
\end{align}
The only difference between the Sj$\ddot{\text{o}}$qvist distance and the Bures distance in Eq. (\ref{db5}) is that the infima are obtained under different conditions. According to Eq. (\ref{Sdis3}), the Sj$\ddot{\text{o}}$qvist distance is invariant under the gauge transformation
diag$(\text{e}^{\text{i}\chi_0},\text{e}^{\text{i}\chi_1},\cdots,\text{e}^{\text{i}\chi_{N-1}})\in$
$\underbrace{\text{U}(1)\times\cdots \times\text{U}(1)}_N$,
which is a subgroup of the U$(N)$ transformation. Therefore, the minimizing condition of Eq. (\ref{Sdis3}) is weaker than that of Eq. (\ref{db5}). This implies
 \begin{align}
 \dif^2_\text{B}(\rho(t),\rho)\le  \dif^2_\text{S}(\rho(t),\rho),
\end{align}
which agrees with the results of Ref. \cite{Alsing23}.

Therefore, the Sj$\ddot{\text{o}}$qvist and Bures distances can be cast into a unified formalism by minimizing the raw distance between nearby purifications with respect to different conditions, making them respectively invariant under the $\text{U}(1)\times\cdots\times \text{U}(1)$ and U$(N)$ gauge transformations. We mainly focus on the Bures distance, given its broad presence in the literature~\cite{PhysRevLett.72.3439,Wiseman_book,Bengtsson_book}. A comparison of the Bures and Sj$\ddot{\text{o}}$qvist metrics of two-level systems is also presented in a recent work~\cite{Alsing24}.

\begin{widetext}

\begin{table}[t]
\centering
\caption{A comparison between the geometries of pure and mixed states.
}\label{table1}
\begin{tabular}{|c|c|c|}\hline
  & Pure state & Mixed state \\ \hline
Total space 
 & $S^{2N-1}$ & $\mathcal{S}_N$  \\ \hline
  Phase space & $CP^{N-1}$ & $\mathcal{D}^N_N$ \\\hline
  Fibration & $S^{2N-1}/\text{U}(1)=CP^{N-1}$ & $\mathcal{S}_N/\text{U}(N)=\mathcal{D}^N_N$ \\\hline
  Connection & Berry connection ($U(1)$ bundle) & Uhlmann connection ($U(N)$ bundle) \\\hline
  Raw distance & $\dif s^2(S^{2N-1})$ & $\dif s^2(\mathcal{S}_N)$ \\\hline
  Gauge-invariant distance & $\dif s^2(CP^{N-1})$ & $\dif s^2_\text{B}(\mathcal{D}^N_N)$ \\\hline
  Relations between distances & $\dif s^2(S^{2N-1})=\dif s^2(CP^{N-1})+(\mi A+\mi\dif g g^\dag)^2$ & $\dif s^2(\mathcal{S}_N)=\dif s^2_\text{B}(\mathcal{D}^N_N)+\text{Tr}\left[\rho (\mi A_\text{U}+\mi\dif UU^\dag)^2\right]$ \\\hline
  Raw metric & $\langle \partial_i\tilde{\psi}|\partial_j\tilde{\psi}\rangle$ & $\langle \partial_\mu W|\partial_\nu W\rangle$\\\hline
  Real part of QGT & \text{Re}$\langle \partial_i\tilde{\psi}|\partial_j\tilde{\psi}\rangle+A_iA_j$ (Fubini-Study) & $g^\text{B}_{\mu\nu}$ (Bures) \\\hline
  Imaginary part of QGT & Berry curvature & Uhlmann form $(=0)$\\\hline
  \end{tabular}
  \end{table}
  \end{widetext}

\subsection{Uhlmann form}
We now turn to the gauge-invariant imaginary part $\sigma^\text{U}_{\mu\nu}$.
A compact expression is obtained by introducing the 2-form
\begin{align}\label{Ufb}
&\sigma^\text{U}=\frac{1}{2}\sigma^\text{U}_{\mu\nu}\dif R^\mu\wedge \dif R^\nu\notag\\
=&\frac{1}{2\mi}\text{Tr}\left[\partial_\mu W^\dag \partial_\nu W-\partial_\mu(W^\dag W\omega_\nu)\right]\dif R^\mu\wedge \dif R^\nu.
\end{align}
We refer to it as the Uhlmann form. To our knowledge, it has not been derived in the literature. Unlike the pure-state case, the Uhlmann form is not proportional to the Uhlmann curvature $F_\text{U}=\dif A_\text{U}+A_\text{U}\wedge A_\text{U}$ with $A_\text{U}$ given by Eq.~(\ref{Au0})
 since the Uhlmann curvature is matrix-valued.
The difference is because $\mathcal{D}^N_N$ does not possess the same features as the K$\ddot{\text{a}}$hler manifold $CP^{N-1}$. Interestingly, when restricted on $\mathcal{D}^N_N$, the Uhlmann form is also independent of the fiber $U$:
\begin{align}\label{Ufbf0}
\sigma^\text{U}=\frac{\mi}{2}\text{Tr}\left[\partial_\mu(\rho A_{\text{U}\nu})\right]\dif R^\mu\wedge \dif R^\nu=\frac{\mi}{2} \text{Tr}[\dif(\rho A_\text{U})].
\end{align}
The proof is summarized in Appendix \ref{appd}. We emphasize that the proof and result can be generalized to situations where the restriction of $\sigma^\text{U}$ on $\mathcal{D}^N_N$ can be relaxed. This is done by replacing $A_\text{U}$ by $\pi^* A_\text{U}$, a one-form on the total space $\mathcal{S}_N$, in the expression above to get
\begin{align}\label{Ufbf0b}
\sigma^\text{U}=\frac{\mi}{2} \text{Tr}[\dif(\rho\pi^* A_\text{U})].
\end{align}
The exterior derivative can be pulled out of the trace if the dimension of the system is finite ($N<\infty$). This leads to vanishing $\sigma^\text{U}$ according to Eq.~(\ref{AU6'}):
\begin{align}\label{Ufbf1b}
\sigma^\text{U}=\frac{\mi}{4}\dif \text{Tr}(\rho\pi^* A_\text{U}+\pi^* A_\text{U}\rho)=\frac{\mi}{4}\dif \text{Tr}[\sqrt{\rho},\dif_{\mathcal{S}_N} \sqrt{\rho}]=0.
\end{align}

What is the physical implication of the vanishing Uhlmann form? In terms of differential forms, Eq.~(\ref{Ufb}) becomes
\begin{align}\label{Ufbb}
\sigma^\text{U}=\frac{1}{2\mi}\dif\text{Tr}(W^\dag\dif W-W^\dag W\omega)=0.
\end{align}
When contracted with a horizontal vector $\tilde{X}=\frac{\dif}{\dif t}$ twice, we have
\begin{align}\label{Ufbd}
\frac{\dif}{\dif t}\text{Tr}(W^\dag\dot{ W})=0
\end{align}
by using $\omega(\tilde{X})=0$.
A horizontal vector generates a parallel-transport condition, which requires $W^\dag\dot{ W}=\dot{W}^\dag W$. By using $\rho=WW^\dag$ and the cyclic property of trace, Eq.~(\ref{Ufbd}) is equivalent to
 \begin{align}\label{Ufbe}
\frac{\dif}{\dif t}\text{Tr}(\dot{ \rho})=0.
\end{align}
For Uhlmann's parallel-transport, this can be recognized as a constraint imposed by the vanishing $\sigma^\text{U}$ because Eq.~(\ref{Ufbe}) is obtained by combing Eq.~(\ref{Ufbd}) and the parallel-transport condition $W^\dag\dot{ W}=\dot{W}^\dag W$. The condition is satisfied by any trace-preserving processes since $\text{Tr}\rho=1$. Therefore, the vanishing Uhlmann form imposes a restriction to rule out parallel transport of open systems where Eq.~\eqref{Ufbe} is violated.
When contracted with a vertical vector $\tilde{X}^V=\frac{\dif}{\dif t}$, Eq.~(\ref{Ufbb}) leads to
\begin{align}\label{Ufbc2}
\frac{\dif}{\dif t}\text{Tr}(W^\dag\dot{ W}-W^\dag W u)=0,
\end{align}
where $u=\omega (\tilde{X}^V) $ is a u$(N)$ generator in the fiber space \cite{10.21468/SciPostPhysCore.6.1.024}. Eq.~(\ref{Ufbc2}) is satisfied by a generic U($N$) transformation on the fiber with $\dot{W}=Wu$, whose solution is $W(t)=W(0)\me^{t u}$. Thus, the vanishing of the gauge-invariant imaginary part associated with the Uhlmann form naturally reflects the physical properties of the Uhlmann parallel transport. Unless exotic processes violating Eq.~\eqref{Ufbe} or \eqref{Ufbc2} are involved, the Uhlmann form remains zero.

We remark that while the pure and mixed states exhibit similar fibrations and geometric structures of local distances, the imaginary part of the QGT of mixed states vanishes identically instead of being proportional to the curvature as the case of pure states. This is because $\mathcal{D}^N_N$ is not necessarily a K$\ddot{\text{a}}$hler manifold. Since $\rho$ is by definition Hermitian, $\mathcal{D}^N_N$ always admits a set of real coordinates not necessarily compatible with an almost complex structure. This can be more clearly understood by noticing that the real dimension of $\mathcal{D}^N_N$ can be an odd number while the dimension of the K$\ddot{\text{a}}$hler manifold $CP^{N-1}$ is $2(N-1)$, which is always even. 

Moreover, the vanishing of the Uhlmann form also highlights the different topologies of pure and mixed states. Here we refer to the U(1) principal bundle of pure states that admits a Berry connection as the Berry bundle, which can be topologically nontrivial. Quantized topological invariants, such as the Chern number, can be found by calculating the characteristic classes of the bundle. In contrast, the Uhlmann bundle is topologically trivial \cite{TDMPRB15,GPbook,OurDPUP,GPbook}, and its characteristic classes must vanish. Since the Uhlmann form is gauge-invariant, its integral over $\mathcal{D}^N_N$ may serve as a topological invariant of the Uhlmann bundle, similar to the Chern number of the Berry bundle. If the integral is non-zero, it would contradict the topological triviality of the Uhlmann bundle. We also remark that the Uhlmann bundle cannot reduce to the Berry bundle as $T\rightarrow0$ because the former requires full-rank density matrices. Since topological properties are defined with respect to the corresponding geometry, the difference between the $T\rightarrow 0$ and $T=0$ cases comes from the change of the underlying bundles. However, there still exists some correspondence between quantities derived from pure and mixed states as $T\rightarrow 0$. which will be discussed in the next section.
Nevertheless, our findings show that $\mathcal{D}^N_N$ already possesses rich geometric properties for mixed states. We summarize our main results by comparing the key points between pure and mixed states in Table \ref{table1}.


\section{Correspondence between pure and mixed states}\label{Sec4}
Since the geometric structures of pure and mixed quantum states are remarkably analogous, one may wonder whether the results of mixed states can reduce to those of pure states as $T\rightarrow 0$ if we consider systems in thermal equilibrium. The fact that $\mathcal{D}^N_N$ of systems in thermal equilibrium cannot become $CP^{N-1}$ of systems in the ground states for $N>1$ because of $\mathcal{D}^N_1=CP^{N-1}$ highlights the challenges connecting the results for pure and mixed states. In contrast, thermodynamic quantities at low temperatures are expected to reduce to their counterparts at zero temperature as $T\rightarrow 0$. Therefore, it is an important task to sort out which geometric quantities of quantum systems approaches their pure-state counterparts as $T\rightarrow 0$.

The fibrations $S^{2N-1}/\text{U}(1)=CP^{N-1}$ and $\mathcal{S}_N/\text{U}(N)=\mathcal{D}^N_N$ lay the foundation for the fiber-bundle descriptions of the Berry and Uhlmann phases, respectively. 
As mentioned previously, the Berry bundle can be topologically nontrivial, whereas the Uhlmann bundle is always trivial.
Consequently, the Uhlmann connection and Uhlmann curvature do not reduce to the Berry connection and Berry curvature as $T\rightarrow 0$ because they belong to fiber bundles with distinct topologies. Nevertheless, it has been conditionally proven that the Uhlmann phase approaches the Berry phase as $T\rightarrow 0$, which is also known as the Uhlmann-Berry correspondence \cite{10.21468/SciPostPhysCore.6.1.024}. One may ask if there exists any other correspondence between the geometric results of pure and mixed states?
Here we show that for systems of dimension $N>1$, 
the Bures metric of mixed states indeed reduces to the Fubini-Study metric of pure states as $T\rightarrow 0$, thereby providing a correspondence.
To our knowledge, there has been no proof of the correspondence between the two metrics. Ref.~\cite{BdSd23} provides an example based on qubits but requires a constant factor to be dropped. Here we provide a general proof showing that the correspondence is exact.

Using the expression (\ref{Bmb}) for the Bures metric, the Bures distance is written as
\begin{align}\label{Bda}
 \dif s^2_\text{B}=\frac{1}{2}\sum_{ij=0}^{N-1}\frac{(\sqrt{\lambda_i}+\sqrt{\lambda_j})^2}{\lambda_i+\lambda_j}|\langle i|\dif\sqrt{\rho}|j\rangle|^2.
\end{align}
Suppose the Hamiltonian of a finite-dimensional quantum system is $\hat{H}$, and the corresponding ground state is $|E_0\rangle\equiv|\tilde{\psi}\rangle$. At temperature $T$ with $\beta=1/T$, the density matrix is given by $\rho=\frac{\me^{-\beta \hat{H}}}{Z}$, where $Z$ is the partition function. In this situation, $\rho$ and $\hat{H}$ share the same set of eigenstates $|i\rangle=|E_i\rangle$, and $\lambda_i=\frac{\me^{-\beta E_i}}{Z}$. From Eq.~(\ref{Bda}), the Bures distance becomes
 \begin{align}\label{Bdb}
 \dif s^2_\text{B}=\sum_{i}\langle i|\dif\sqrt{\rho}|i\rangle^2+\frac{1}{2}\sum_{i\neq j}\frac{(\sqrt{\lambda_i}+\sqrt{\lambda_j})^2}{\lambda_i+\lambda_j}|\langle i|\dif\sqrt{\rho}|j\rangle|^2.
\end{align}
For simplicity, we focus on the situation with no energy degeneracy. Let $E_0<E_1<\cdots< E_{N-1}$, then
\begin{align}\label{lm}
\lim_{T\rightarrow 0}\frac{\lambda_{i}}{\lambda_j}=\lim_{\beta\rightarrow \infty}\me^{-\beta(E_i-E_j)}=0, \quad \text{if } i>j.
\end{align}
For $i\neq j$, we set $\lambda_\text{min}=\min\{\lambda_i,\lambda_j\}$ and $\lambda_\text{max}=\max\{\lambda_i,\lambda_j\}$. This implies
\begin{align}
\lim_{T\rightarrow 0}\frac{\left(\sqrt{\lambda_i}+\sqrt{\lambda_j}\right)^2}{\lambda_i+\lambda_j}=
\lim_{T\rightarrow 0}\frac{\left(1+\sqrt{\frac{\lambda_\text{min}}{\lambda_\text{max}}}\right)^2}{1+\frac{\lambda_\text{min}}{\lambda_\text{max}}}=1.
\end{align}
Thus, Eq.~(\ref{Bdb}) reduces to
 \begin{align}\label{Bdc}
\lim_{T\rightarrow 0} \dif s^2_\text{B}
&=\frac{1}{2}\sum_{i}\langle i|\dif\sqrt{\rho}|i\rangle^2+\frac{1}{2}\text{Tr}(\dif\sqrt{\rho}\dif\sqrt{\rho}).
\end{align}
The normalized eigenstate $|i\rangle$ satisfies $\langle i|\dif i\rangle+\langle\dif i| i\rangle=0$, resulting in $\langle i|\dif\sqrt{\rho}|i\rangle=\dif \sqrt{\lambda_i}$. A straightforward evaluation shows (see Appendix \ref{appf} for details)
 \begin{align}\label{Bdd0}
\lim_{T\rightarrow 0} \dif s^2_\text{B}&=\sum_i(\dif\sqrt{\lambda_i})^2+\sum_i\lambda_i\langle \dif i|\dif i\rangle\notag\\&+\sum_{ij}\sqrt{\lambda_i\lambda_j}\langle i|\dif j\rangle\langle  j|\dif i\rangle.
\end{align}
As $T\rightarrow 0$, $\lambda_0\rightarrow 1$ and $\lambda_{i>1}\rightarrow 0$, both of which are independent of the evolution parameters. As a consequence, $\dif \sqrt{\lambda_i}$=0.
This leads to
\begin{align}\label{Bdf}
&\lim_{T\rightarrow 0} \dif s^2_\text{B}=\lambda_0(\langle \dif \tilde{\psi}|\dif\tilde{\psi}\rangle+\langle \tilde{\psi}|\dif \tilde{\psi}\rangle\langle  \tilde{\psi}|\dif \tilde{\psi}\rangle)\notag\\+&\lambda_0\sum_i\frac{\lambda_i}{\lambda_0}\langle \dif i|\dif i\rangle-\lambda_0\sum_{i>0\text{ or } j>0}\sqrt{\frac{\lambda_i\lambda_j}{\lambda_0^2}}|\langle i|\dif j\rangle|^2.
\end{align}
Using Eq.~\eqref{lm}
and $\lim_{T\rightarrow 0}\lambda_0\rightarrow 1$ because $\lambda_0+\cdots +\lambda_{N-1}=1$, the correspondence is established:
\begin{align}\label{Bdg}
\lim_{T\rightarrow 0} \dif s^2_\text{B}=\langle \dif \tilde{\psi}|\dif\tilde{\psi}\rangle+\langle \tilde{\psi}|\dif \tilde{\psi}\rangle^2=\dif s^2_\text{FS},
\end{align}
according to Eq.~(\ref{QGT0}). Therefore, the Bures metric reduces to the Fubini-Study metric in the zero temperature limit for finite-dimensional quantum systems in thermal equilibrium.

The relation between the Bures and Fubini-Study metrics as temperature approaches zero builds a correspondence between the real parts of the QGTs of pure and mixed states.
On the other hand, there is no such correspondence between the associated imaginary parts of the QGTs of the pure and mixed states. As shown in the previous discussions, the imaginary part of the mixed-state QGT is the Uhlmann form, which is identically zero, while the imaginary part of the pure-state QGT is proportional to the Berry curvature.

For the simplest situation with $N=1$, the density matrix is of rank 1 and actually describes a pure state: $\rho=|\psi\rangle\langle \psi|$. Let $\rho(t)=|\psi(t)\rangle\langle \psi(t)|$. Using $\rho=\sqrt{\rho}$, the Bures distance between $\rho$ and $\rho(t)$ is
 \begin{align}\label{Bd2}
\dif^2_\textrm{B}(\rho,\rho(t))=2-2|\langle \psi|\psi(t)\rangle|
\end{align}
according to Eq.~(\ref{db3}). Indeed, this reduces to the Fubini-Study distance (\ref{dFS}) for pure states. However, this reduction intrinsically differs from the more general correspondence because the Bures metric is trivial in this case:  Since $\lambda_i=\lambda_j=1$ and $|i\rangle=|j\rangle=|\psi\rangle$,
 \begin{align}\label{Bm2}
\langle i|\partial_\mu\rho|j\rangle=\langle\psi|\left(|\partial_\mu\psi\rangle\langle\psi|+|\psi\rangle\langle\partial_\mu\psi|\right)|\psi\rangle=0
\end{align}
by using $\langle\psi|\psi\rangle=1$. Thus, $g_{\mu\nu}^\text{B}=0$ according to Eq.~(\ref{Bm}). This is because $\dim \mathcal{D}^1_1=1^2-1=0$, representing a single point as we have pointed out before.

\section{Examples and Implications}\label{Sec5}
We will present concrete examples showing the correspondence between the Bures and Fubini-Study metrics.
\subsection{Generic result for two-level systems}
For a two-level system with $N=2$, the expression (\ref{A11}) of the Bures distance can be further simplified. As the generators $T_i$ in Eq.~(\ref{dmd1b}) reduce to the Pauli matrices, the density matrix accordingly takes the form $\rho=\frac{1}{2}+\mathbf{a}(\mathbf{R})\cdot\boldsymbol{\sigma}$. In this case, the Bures distance reduces to (see Appendix \ref{appa} for details)
 \begin{align}\label{A12}
\dif^2_\text{B}(\rho,\rho+\mathrm{d}\rho)&=\frac{1}{2}\textrm{Tr}(\mathrm{d}\rho)^2+(\mathrm{d}\sqrt{\det\rho})^2\notag\\
&=\mathrm{d}\mathbf{a}\cdot \mathrm{d}\mathbf{a}+\frac{(\mathbf{a}\cdot \mathrm{d}\mathbf{a})^2}{b^2},
\end{align}
where $b=\sqrt{\det\rho}=\sqrt{\frac{1}{4}-\mathbf{a}^2}$. Thus, the corresponding Bures metric is
\begin{align}\label{Bm3}
g^\text{B}_{\mu\nu}=\frac{\partial\mathbf{a}}{\partial R^\mu}\cdot\frac{\partial\mathbf{a}}{\partial R^\nu}+\frac{1}{\frac{1}{4}-\mathbf{a}^2}\mathbf{a}\cdot\frac{\partial\mathbf{a}}{\partial R^\mu}\mathbf{a}\cdot\frac{\partial\mathbf{a}}{\partial R^\nu}.
\end{align}
To compare with the pure-state results, we set $\mathbf{a}=\frac{\mathbf{x}}{2}$ for convenience
and let $r=|\mathbf{x}|$ and $\mathbf{n}=\frac{\mathbf{x}}{r}$. Using $\mathbf{n}\cdot\dif\mathbf{n}=0$, the Bures distance takes the form \cite{PhysRevB.97.094110}
\begin{align}\label{A13}
\dif^2_\text{B}(\rho,\rho+\mathrm{d}\rho)=\frac{1}{4}\left(\frac{\dif r^2}{1-r^2}+r^2\dif\mathbf{n}\cdot\dif\mathbf{n}\right).
\end{align}
As a comparison, Eq.~(\ref{dHSrho2}) indicates that the Hilbert-Schmidt distance is
\begin{align}\label{A13b}
\dif^2_\text{HS}(\rho,\rho+\mathrm{d}\rho)=\frac{1}{4}\left(\dif r^2+r^2\dif\mathbf{n}\cdot\dif\mathbf{n}\right),
\end{align}
which has the familiar Euclidean-distance form.

When $r\rightarrow 1$, $\rho$ reduces to the pure-state density matrix. However, it appears that the Bures distance diverges as $r\rightarrow 1$.
This seemingly singular behavior is actually spurious because $\dif r$ also approaches 0 as $r\rightarrow 1$. Since $0\le r < 1$, a change of variable $r=\cos u$ leads Eq.~(\ref{A13}) to
\begin{align}\label{A14}
\dif^2_\text{B}(\rho,\rho+\mathrm{d}\rho)=\frac{1}{4}\left(\dif u^2+\cos^2 u\dif\mathbf{n}\cdot\dif\mathbf{n}\right).
\end{align}
It reduces to $\dif^2_\text{B}(\rho,\rho+\mathrm{d}\rho)=\frac{1}{4}\dif\mathbf{n}\cdot\dif\mathbf{n}$ as $r\rightarrow 1$. Introducing $\mathbf{n}=(\sin\theta\cos\phi,\sin\theta\sin\phi,\cos\theta)^T$, the Bures distance further reduces to
\begin{align}\label{A15}
\dif^2_\text{B}(\rho,\rho+\mathrm{d}\rho)=\frac{1}{4}\left(\dif\theta^2+\sin\theta^2\dif\phi^2\right)
\end{align}
as $T\rightarrow 0$, which agrees with the Fubini-Study distance for pure states shown in Eq.~(\ref{ds2CP1}).

\begin{figure}[t]
\centering
\includegraphics[width=2.2in]{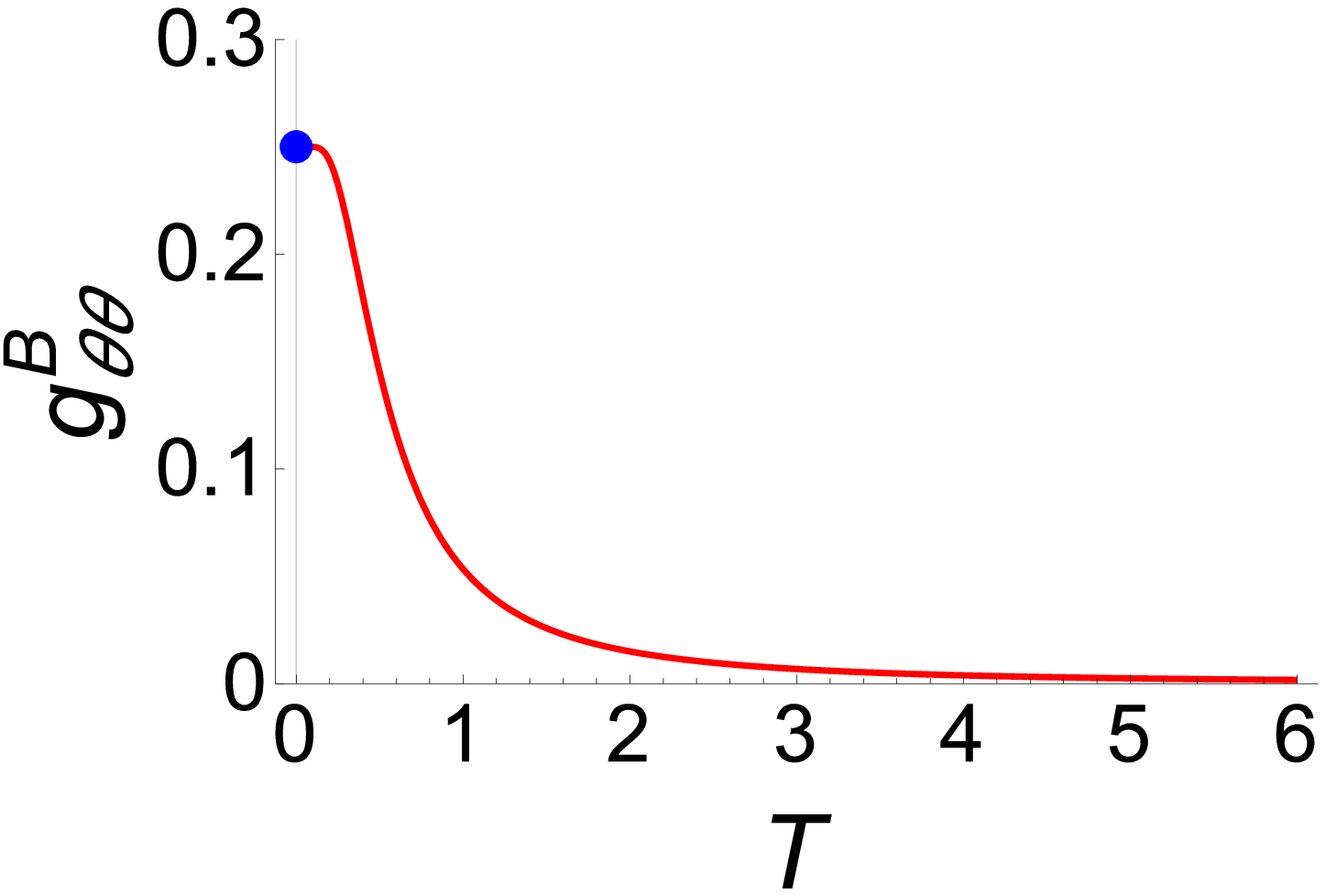}
\includegraphics[width=2.2in]{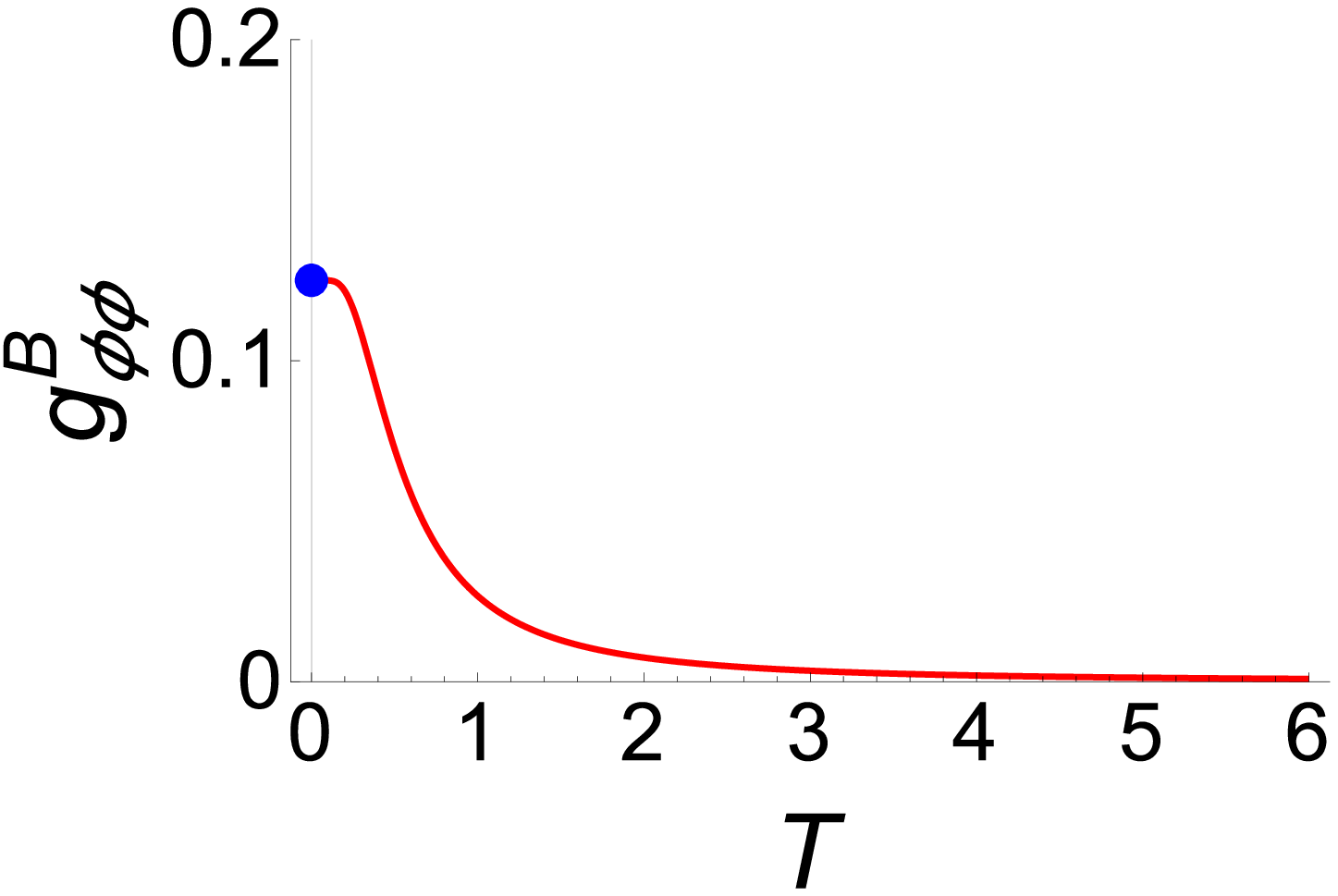}
\caption{Bures metric of the spin-$\frac{1}{2}$ paramagnet as a function of temperature: (Top panel) $g^\text{B}_{\theta\theta}$, (Bottom panel) $g^\text{B}_{\phi\phi}$ at $\theta=\frac{\pi}{4}$. 
The dots at $T=0$ show the corresponding values from the Fubini-Study metric.
}
\label{Fig1}
\end{figure}

\subsection{Spin-$\frac{1}{2}$ system}
We now study a two-level system that is physically equivalent to many experimentally realized systems at zero temperature
\cite{Gianfrate2020,10.1093/nsr/nwz193,PhysRevLett.122.210401}.
We consider an ensemble ensemble of spin-$\frac{1}{2}$ paramagnets influenced by an external magnetic field $\mathbf{B}$ with fixed magnitude $B=|\mathbf{B}|$ and described by the Hamiltonian
\begin{equation}
\hat{H}=\omega_{0} \hat{\bB}\cdot \frac{\boldsymbol{\sigma}}{2}.
\end{equation}
Here $\omega_{0}$ is the Larmor frequency, and $ \hat{\bB}=\mathbf{B}/B$. The orientation of $\mathbf{B}$ can be controlled externally via the angles $\theta$ and $\phi$: $\mathbf{B}=B\left (
\sin\theta \cos\phi,
\sin\theta \sin\phi,
\cos\theta
\right )^T$. At temperature $T$ with $\beta=\frac{1}{T}$, the thermal equilibrium density matrix is $\rho(T)=\frac{\me^{-\beta\hat{H}}}{\text{Tr}\me^{-\beta\hat{H}}}=\frac{1}{2}[1-\tanh(\frac{\beta \omega_{0}}{2})\hat{\bB}\cdot \boldsymbol{\sigma}]$. Thus, the Bloch vector is $\mathbf{a}=-\frac{1}{2}\tanh(\frac{\beta \omega_{0}}{2})\hat{\bB}$, whose magnitude depends on temperature.

Using Eq.~(\ref{Bm3}), the Bures metrics are given by
\begin{align}
g^\text{B}_{\theta\theta}&=\frac{1}{4}\tanh^2\left(-\frac{\beta \omega_{0}}{2}\right), \notag \\
g^\text{B}_{\phi\phi}&=\frac{1}{4}\tanh^2\left(-\frac{\beta \omega_{0}}{2}\right)\sin^2\theta, \notag \\
g^\text{B}_{\theta\phi}&=0,
\end{align}
which are proportional to the ordinary metric of $S^2$ with a temperature-dependent scaling factor $\frac{1}{4}\tanh^2(-\frac{\beta \omega_{0}}{2})$. This becomes clearer by examining the Bures distance via Eq.~(\ref{A13}) at fixed temperature:
\begin{align}
\dif^2_\text{B}(\rho,\rho+\dif\rho)=\frac{1}{4}\tanh^2\left(-\frac{\beta \omega_{0}}{2}\right)\left(\dif\theta^2+\sin^2\theta\dif\phi^2\right).\notag
\end{align}
Therefore, we only concentrate on the variation of the Bures metric with temperature since the metric of $S^2$ is trivial. Figure \ref{Fig1} presents the dependence of $g^\text{B}_{\theta\theta}$ and $g^\text{B}_{\phi\phi}$ on $T$ with $\theta=\frac{\pi}{4}$. As $T\rightarrow 0$, they reduce to the Fubini-Study  metric (\ref{ds2CP1}) since $g^\text{B}_{\theta\theta}(T\rightarrow 0)=\frac{1}{4}$, $g^\text{B}_{\phi\phi}(T\rightarrow 0)=\frac{1}{4}\sin^2\theta$. This confirms our previous assertion about the correspondence between the Bures and Fubini-Study metrics since the eigenvalues of $\rho$ are constants due to the fact that $B=|\mathbf{B}|$ is fixed.
As $T\rightarrow \infty$, $g^\text{B}_{\theta\theta}=g^\text{B}_{\phi\phi}=0$. This is reasonable since the density matrix is $\rho=\frac{1}{2}$ in the infinite-temperature limit. The Bloch vector is $\mathbf{a}=\mathbf{0}$, corresponding to the origin of the Bloch ball, whose neighborhood also collapses to the origin as $T\rightarrow +\infty$ and loses its local structure. We emphasize that the results are relevant to experimentally realizable systems~\cite{Gianfrate2020,10.1093/nsr/nwz193,PhysRevLett.122.210401} and may guide future measurements of the QGT at finite temperatures.

\begin{figure}[t]
\centering
\includegraphics[width=2.2in]{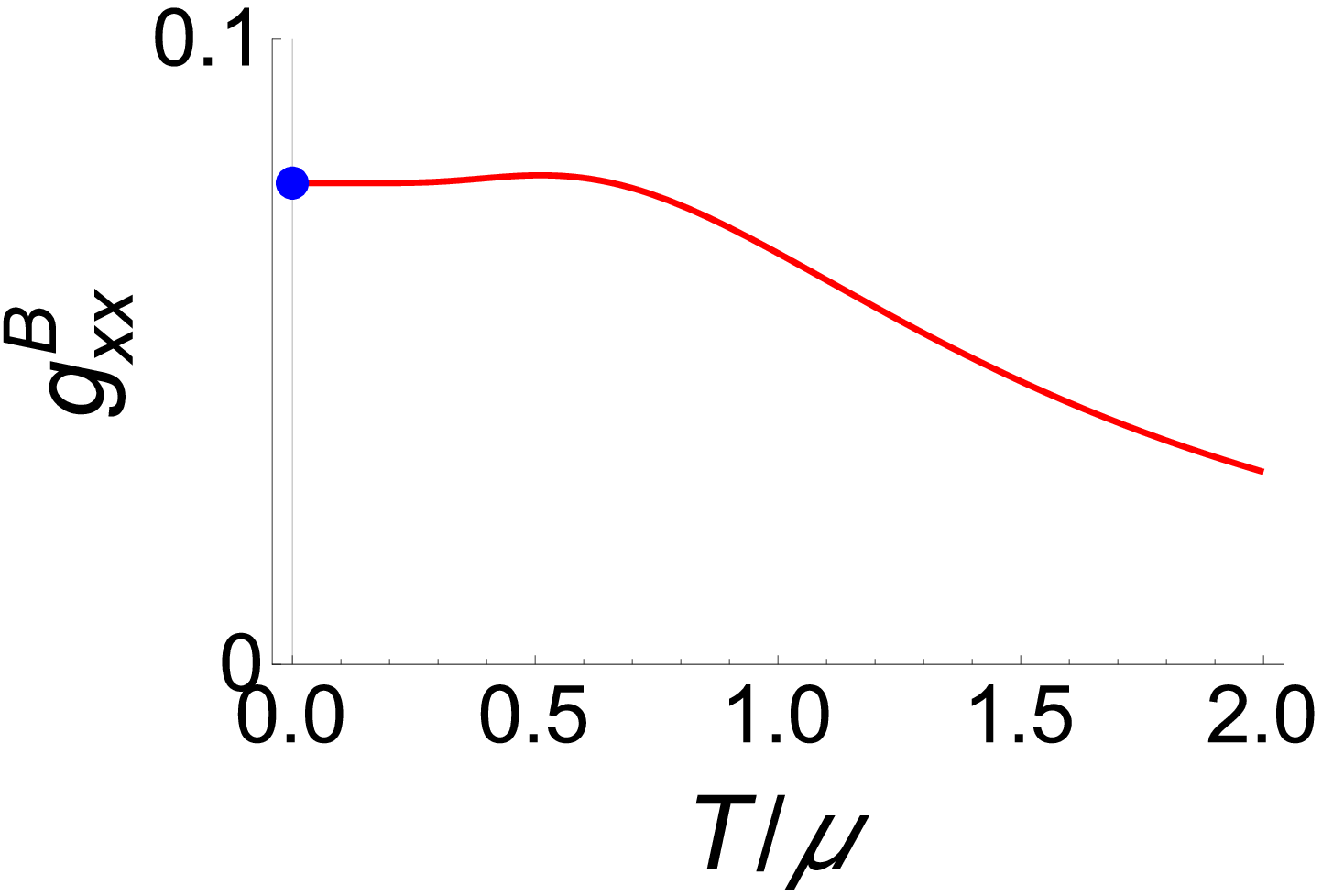}
\includegraphics[width=2.4in]{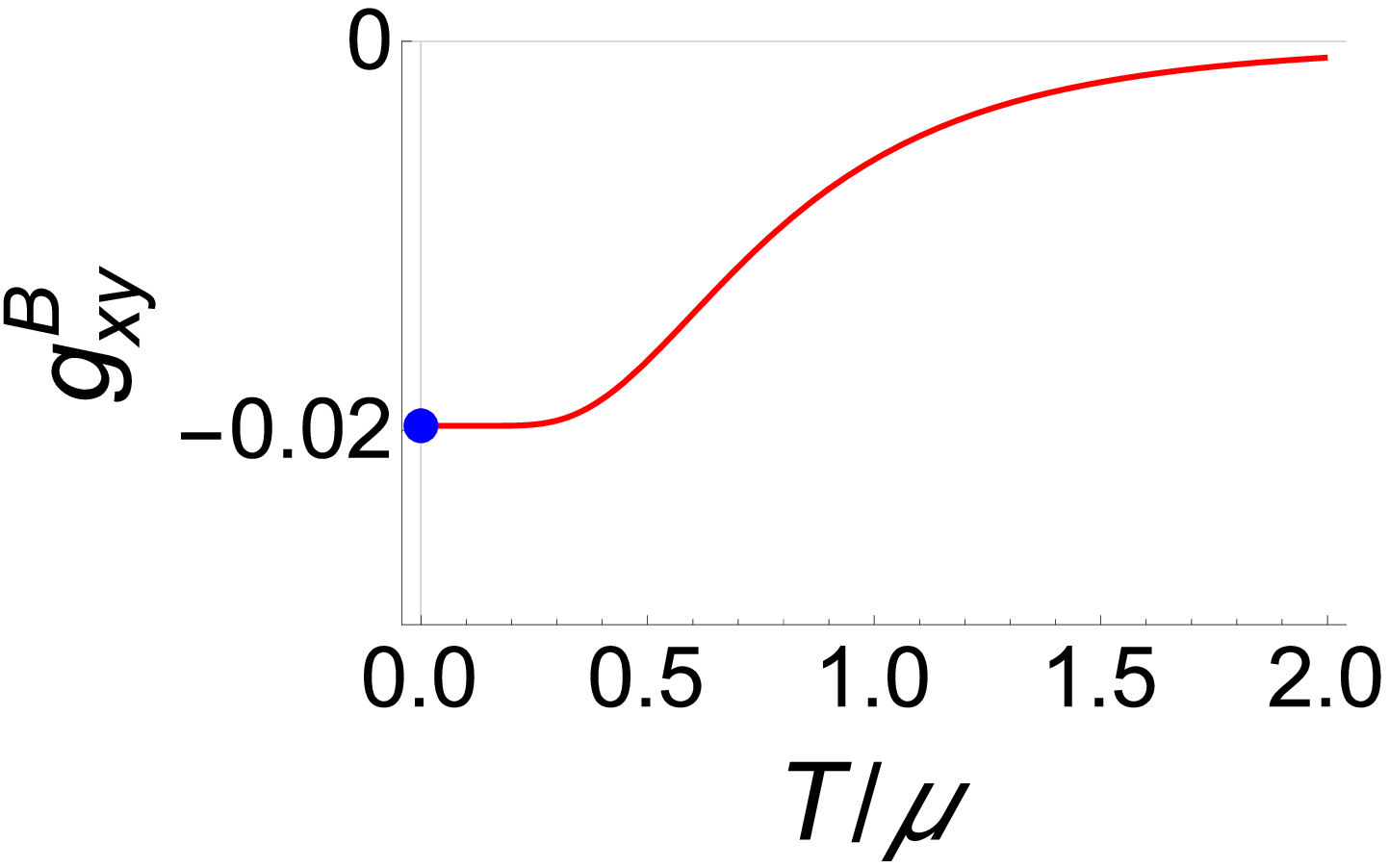}
\caption{Bures metric of the toy model as a function of temperature, (Up) $g^\text{B}_{xx}$, (Bottom) $g^\text{B}_{xy}$, where $k_x=k_y=1.2\pi$ and $\mu=1.0$.
The dots at $T=0$ show the corresponding values from the Fubini-Study metric.
}
\label{Fig2}
\end{figure}

\subsection{2D two-band model}
In the first example, the Bures metric
has zero off-diagonal terms and the eigenvalues of the density matrix are independent of the evolution parameters.  
Here we show another example where the QGT has different behavior. We consider a two-band model inspired by Refs.~ \cite{Bernevigbook,PhysRevB.74.085308,Volovik99,ReadGreen00,Ivanov01,RevModPhys.80.1083}. In momentum space, the Hamiltonian is given by
\begin{align}\label{H2DTS}
\hat{H}_\mathbf{k}=\sin k_{x}\sigma_{x}+\sin k_{y}\sigma_{y}+\mu\sigma_z,
\end{align}
where $\mu>0$ and $\mathbf{k}=(k_x,k_y)^T$ is the 2D crystal momentum. We assume the parameters are independent of temperature. The density matrix in thermal equilibrium is given by
\begin{align}
\rho_{\bk}(T)=\frac{\me^{-\beta \hat{H}_{\bk}}}{\text{Tr}(\me^{-\beta \hat{H}_{\bk}})}
=\frac{1}{2}\left[1-\tanh\left(\frac{\beta \Delta_{\bk}}{2}\right)\hat{\mathbf{n}}_{\bk}\cdot \boldsymbol{\sigma}\right], \notag
\end{align}
where $\hat{\mathbf{n}}_{\bk}=\frac{2}{\Delta_{\bk}}(\sin k_{x},\sin k_{y},\mu)$ with $\Delta_{\bk}=2\sqrt{\sin^2k_{y}+\sin^2k_{x}+\mu^2}$ being the gap in the spectrum of $\hat{H}_\bk$.
In this case, the eigenvalues of the density matrix depend on $\mathbf{k}$ and give nontrivial contributions to the Bures metric.
The Bloch vector is $\mathbf{a}_\bk=-\frac{1}{2}\tanh\left(\frac{\beta \Delta_{\bk}}{2}\right)\hat{\mathbf{n}}_{\bk}$ such that $\rho_\bk=\frac{1}{2}(1+\mathbf{a}_\bk\cdot \boldsymbol{\sigma})$.
The Bures metric is then given by
\begin{equation}\label{TDTSBD}
g^\text{B}_{ij}=\frac{\partial\mathbf{a}_{\bk}}{\partial k_{i}}\cdot\frac{\partial\mathbf{a}_{\bk}}{\partial k_{j}}+\frac{1}{\frac{1}{4}-|\mathbf{a}_{\bk}|^2}\mathbf{a}_{\bk}\cdot\frac{\partial\mathbf{a}_{\bk}}{\partial k_{i}}\mathbf{a}_{\bk}\cdot\frac{\partial\mathbf{a}_{\bk}}{\partial k_{j}},
\end{equation}
for $i,j=k_x,k_y$.
A straightforward evaluation shows
\begin{widetext}
\begin{align}
g^\text{B}_{xx}&=\Bigg[\frac{\beta^2 }{4\cosh^2(\frac{\beta \Delta_{\bk}}{2})}-\frac{\tanh^2(\frac{\beta \Delta_{\bk}}{2})}{\Delta^2_{\bk}} \Bigg]\frac{4\sin^2 k_{x}\cos^2 k_{x}}{\Delta^2_{\bk}}  +\frac{\tanh^2(\frac{\beta \Delta_{\bk}}{2})}{\Delta^2_{\bk}}\cos^2 k_{x}, \notag \\
g^\text{B}_{yy}&=\Bigg[\frac{\beta^2 }{4\cosh^2(\frac{\beta \Delta_{\bk}}{2})}-\frac{\tanh^2(\frac{\beta \Delta_{\bk}}{2})}{\Delta^2_{\bk}} \Bigg]\frac{4\sin^2 k_{y}\cos^2 k_{y}}{\Delta^2_{\bk}}  +\frac{\tanh^2(\frac{\beta \Delta_{\bk}}{2})}{\Delta^2_{\bk}}\cos^2 k_{y}, \notag \\
g^\text{B}_{xy}&=\Bigg[\frac{\beta^2 }{4\cosh^2(\frac{\beta \Delta_{\bk}}{2})}-\frac{\tanh^2(\frac{\beta \Delta_{\bk}}{2})}{\Delta^2_{\bk}} \Bigg]\frac{4\sin k_{x}\cos k_{x}\sin k_{y}\cos k_{y}}{\Delta^2_{\bk}} .
\end{align}
\end{widetext}

This example differs qualitatively from the spin-$\frac{1}{2}$ paramegnet because the eigenvalues of $\rho$ now depends on the Bloch momentum.
The QGT has nonzero off-diagonal term, and the Bures distance is not simply that of $S^2$ up to a scaling factor. Moreover, $g^\text{B}_{xx}$ and $g^\text{B}_{yy}$ are swapped under $(k_x,k_y)\rightarrow (k_y,k_x)$ and $(k_x,k_y)\rightarrow (-k_y,k_x)$ while $g^\text{B}_{xy}$ is even under the first interchange but odd under the second~\cite{Referee}. Thus, the reduction of the Bures metric to the Fubini-Study metric when $T\rightarrow 0$ is not as trivial as the prior example, despite the guarantee from the aforementioned proof. A straightforward calculation of the Fubini-Study metric of this case yields
\begin{align}
& g^\text{FS}_{x x}(k_x, k_y)=\frac{\cos ^2k_x(\sin ^2k_y+\mu^2)}{(\sin ^2k_x+\sin ^2k_y+\mu^2)^2}, \notag\\
& g^\text{FS}_{y y}(k_x, k_y)=\frac{\cos ^2k_y(\sin ^2k_x+\mu^2)}{(\sin ^2k_x+\sin ^2k_y+\mu^2)^2}, \notag\\
& g^\text{FS}_{x y}(k_x, k_y)=-\frac{1}{4} \frac{\sin 2 k_x \sin 2 k_y}{(\sin ^2k_x+\sin ^2k_y+\mu^2)^2}.
\end{align}
By taking the proper limit, it can be verified that $g^\text{B}_{ij}$ indeed reduces to $g^\text{FS}_{ij}$ as $T\rightarrow 0$.

\begin{figure}[t]
\centering
\includegraphics[width=1.82in]{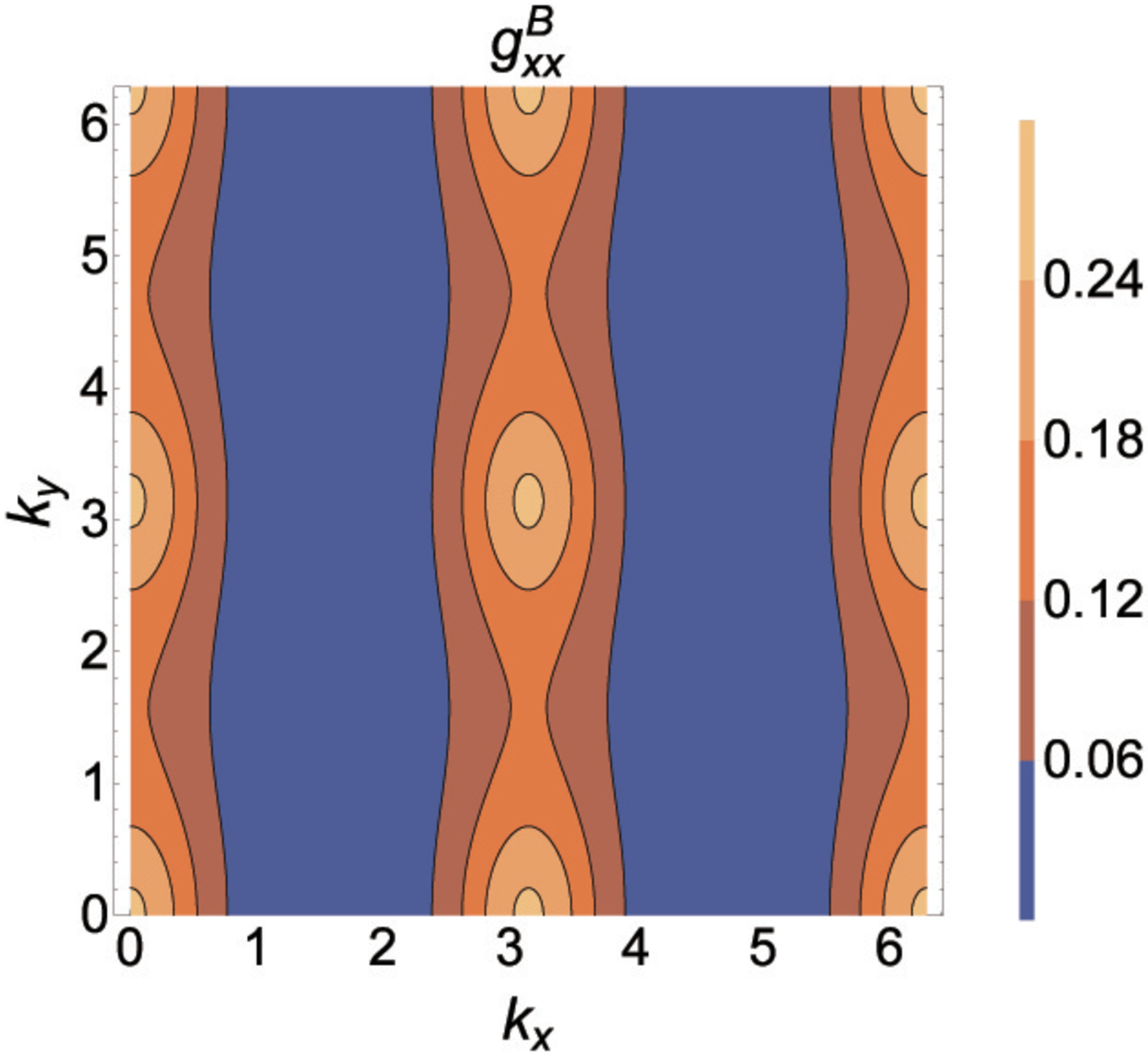}\includegraphics[width=1.55in]{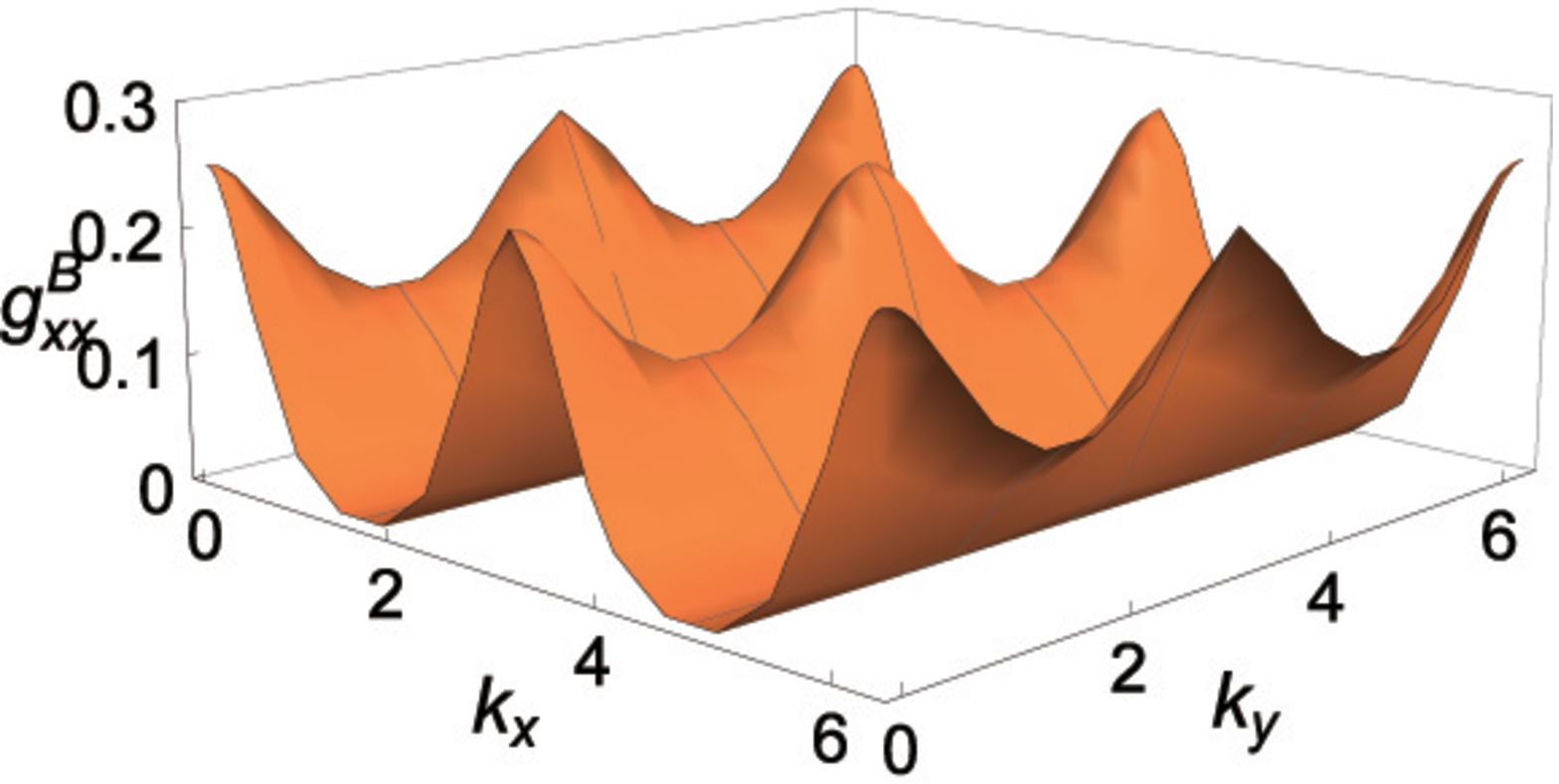}\\
\includegraphics[width=1.82in]{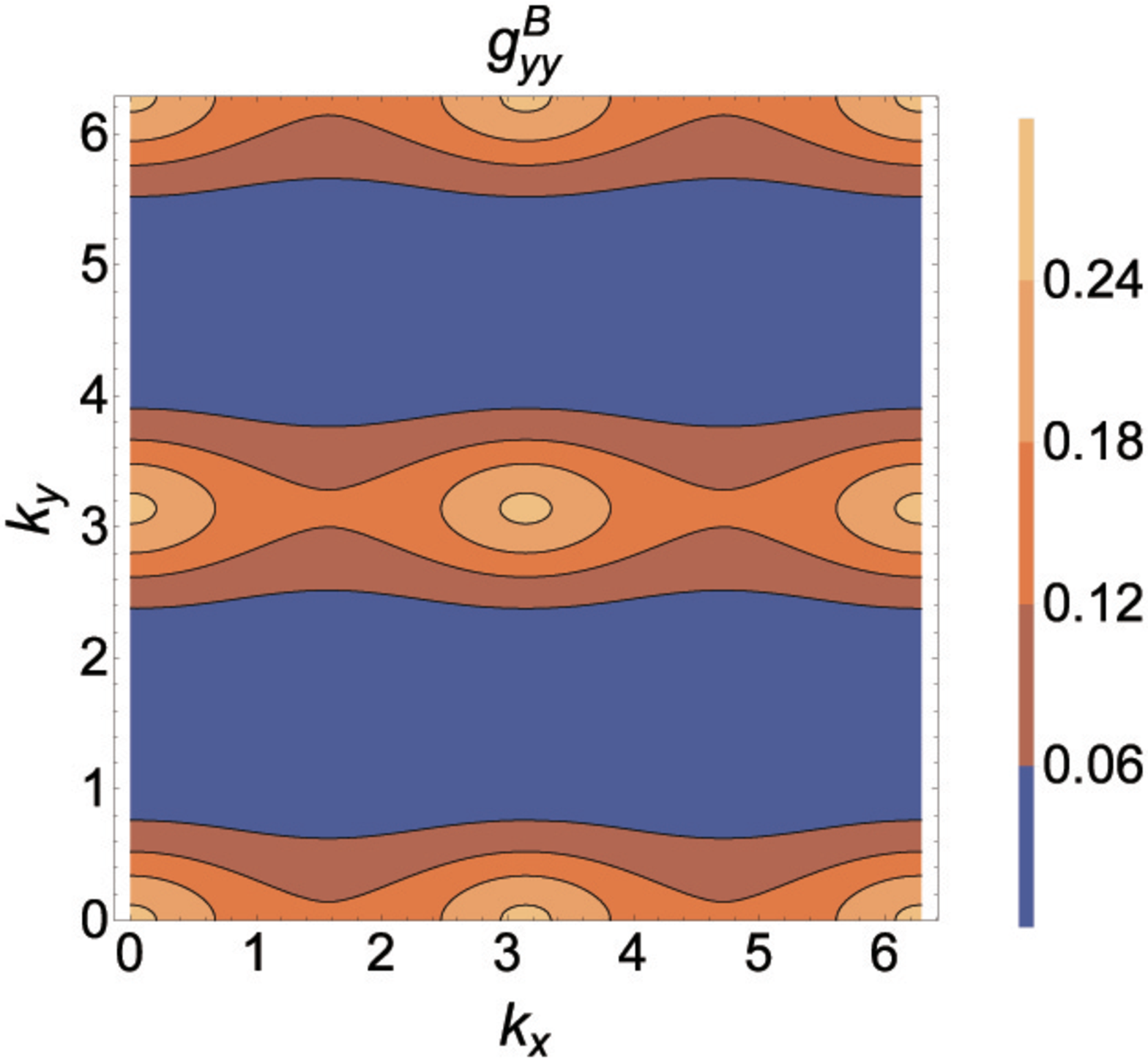}\includegraphics[width=1.55in]{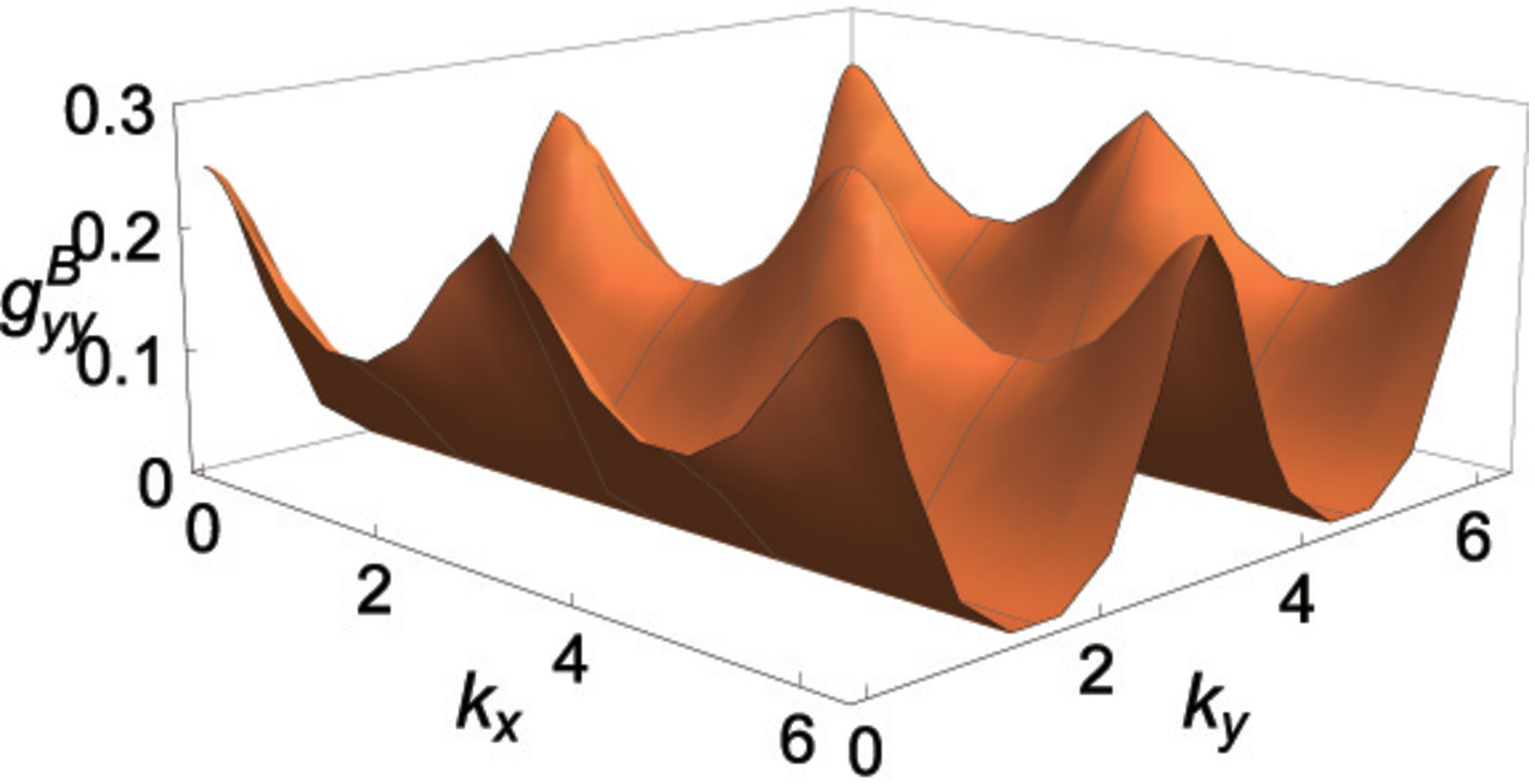}\\
\includegraphics[width=1.82in]{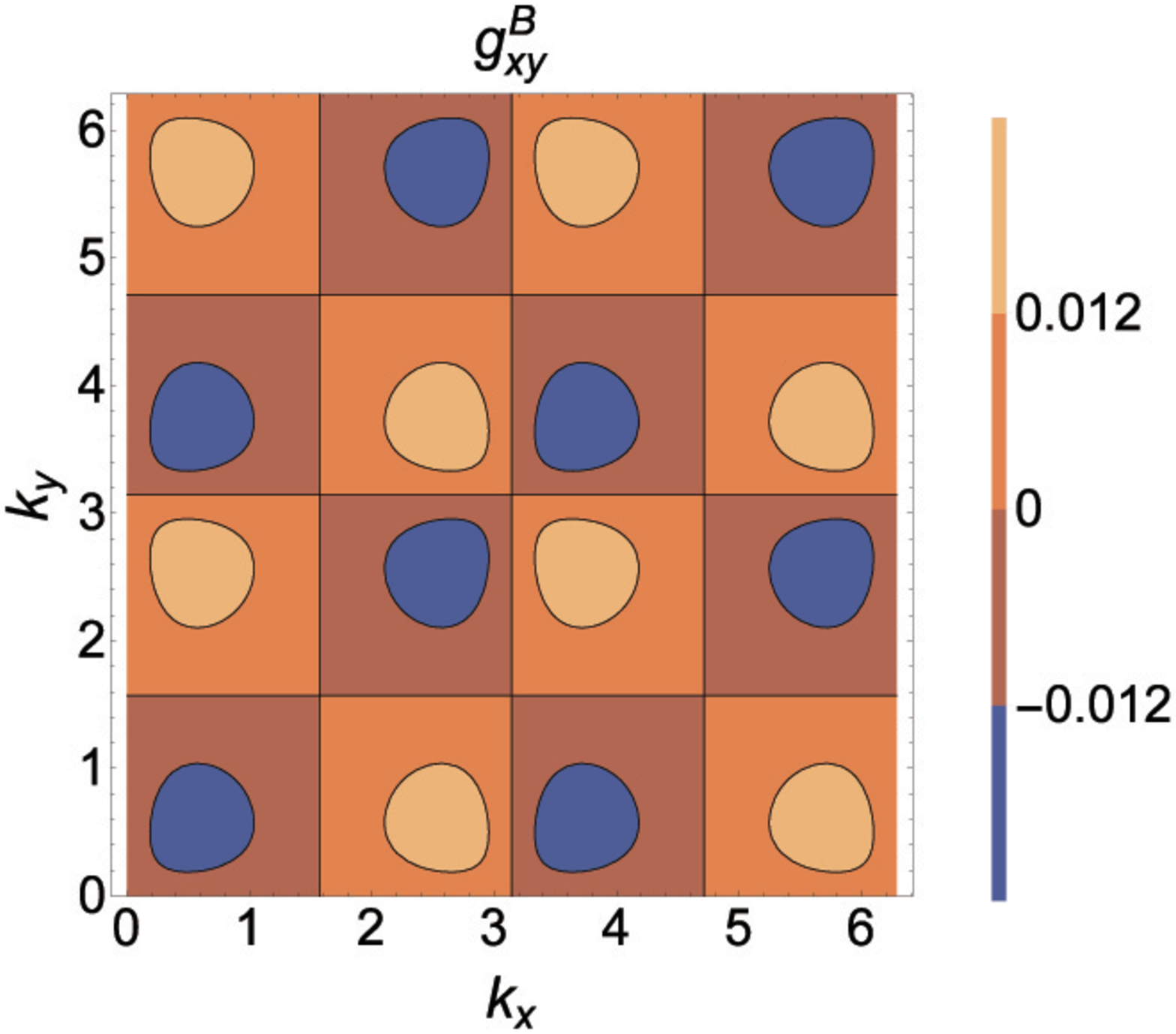}\includegraphics[width=1.55in]{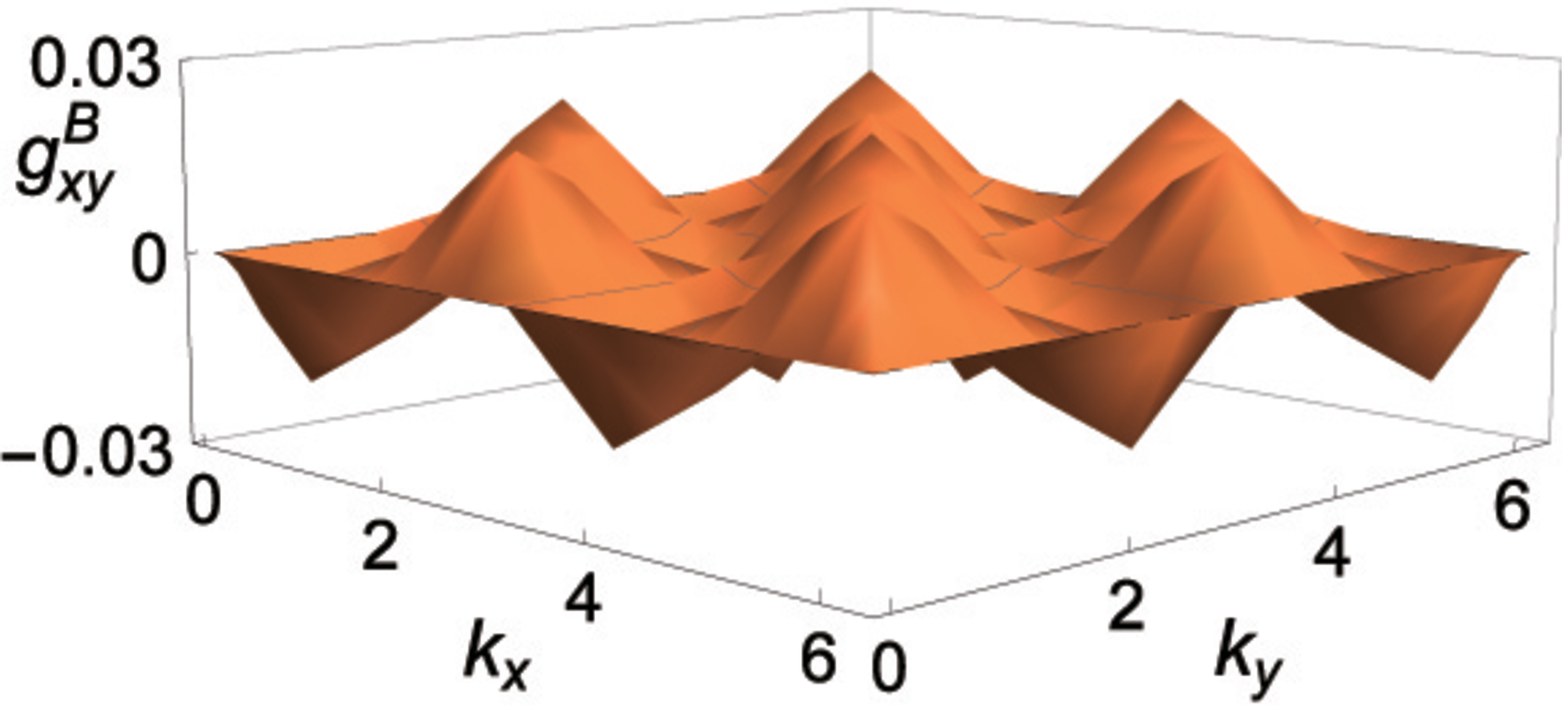}
\caption{The Bures metric of the model~\eqref{H2DTS} in the first Brillouin zone  at $T=0.1\mu$. The left (right) column shows the contour (3D) plots.  }
\label{Fig3}
\end{figure}

We plot the component $g^\text{B}_{xx}$ of the Bures metric as a function of $T$ in Fig.~\ref{Fig2} by setting $k_x=k_y=1.2\pi$, where $g^\text{B}_{yy}$ is not shown since its behavior is the same as $g^\text{B}_{xx}$ up to the $k_x\leftrightarrow k_y$ symmetry. The off-diagonal term $g^\text{B}_{xy}$ is negative in this case. However, this is possible for a curved manifold and does not affect the positivity of the Bures distance $\dif s^2_\text{B}(\mathcal{D}^N_N)$.
Unlike the spin-$\frac{1}{2}$ system, $g^\text{B}_{xx}(k_x=1.2\pi,k_y=1.2\pi)$ reaches its maximum at finite temperature, showing the complicated dependence on temperature of the Bures metric for this model. If $T\rightarrow +\infty$, all components of the Bures metric decay to zero as expected due to the triviality of the density matrix at infinite temperature.

Figure \ref{Fig3} shows the behavior of the Bures metrics of the model in the first Brillouin zone at $T=0.1\mu$.
The left column shows the contour plots for $g^\text{B}_{xx}$, $g^\text{B}_{yy}$, and $g^\text{B}_{xy}$, respectively, and the right column shows the corresponding 3D plots. The results clearly respect the aforementioned symmetries. $g^\text{B}_{xx}$ and $g^\text{B}_{yy}$ both have a regular peak-array structure and are always positive, but $g^\text{B}_{xy}$ possesses both peaks and basins and may also be negative.
The peaks of $g^\text{B}_{xx}$ and $g^\text{B}_{yy}$ appear at $\mathbf{k}_c=(n\pi,m\pi)^T$ with $n,m=0,1,2$, where the energy gap takes its minimum value $\Delta_{min}=2\mu$. These points ($\mathbf{k}_c$) are also saddle points of the contour plot of $g^\text{B}_{xy}$, as depicted in Fig.~\ref{Fig3}. Moreover, since the off-diagonal term $g^\text{B}_{xy}(\mathbf{k}_c)=0$ and the diagonal term $g^\text{B}_{xx,yy}(\mathbf{k}_c)$ take the maximal values, the Bures distance reaches its local maximum at $\mathbf{k}_c$.
The rich structures of the Bures metric will inspire future research on the QGT of mixed quantum states.

\subsection{Implications}
\subsubsection{Physical relevance}
Understanding the geometry behind the QGT of mixed states will allow us to advance the investigations of quantum systems beyond their ground-state properties. For example, previous studies of the Bures distance in entanglement~\cite{PhysRevA.61.064301,PhysRevA.68.062309,PhysRevLett.106.190502}, quantum discord~\cite{Spehner_2013}, quantum criticality~\cite{PhysRevA.75.032109,PhysRevA.76.062318}, Gaussian states~\cite{PhysRevA.93.052330}, quantum parameter estimation~\cite{PhysRevA.88.040102}, and comparisons with other quantum distances~\cite{PhysRevA.84.032120,Alsing23} can now be phrased in a unified framework by the geometry of mixed quantum states and their QGTs.

On the other hand, we have shown that the QGT of mixed states only manifest itself in the Bures distance, as the imaginary part given by the Uhlmann form vanishes for regular processes. Therefore, the applications of the Bures distance mentioned above also provide experimentally testable means for checking the mixed-state QGT in various quantum systems.
The Bures distance is also closely related to the Uhlmann fidelity \cite{HubnerPLA93,Uhlmann95}, which can be expressed in terms of the partition function \cite{PhysRevA.75.032109}. If the system is driven by a Zeeman-like term, the fidelity susceptibility can be inferred from the
magnetic susceptibility \cite{PhysRevE.76.022101}, thereby connecting the QGT with an experimentally measurable quantity.
Furthermore, the vanishing of the gauge-invariant imaginary part of the QGT of mixed states provides a no-go theorem, in contrast to the case of the pure-state QGT, whose imaginary part is the Berry curvature with measurable consequences.

\subsubsection{Challenges beyond full-rank density matrices}
One important reason behind the choice of full-rank density matrix in the construction of the Uhlmann bundle is to ensure the uniqueness of the polar decomposition of the amplitude, i.e., $W=\sqrt{\rho}U$. The full-rank density matrices already cover a great portion of mixed states, including systems at finite temperatures. Going beyond the full-rank density matrices will require a different construction of the underlying fiber bundle and the associated local geometry.
As an oversimplified example, we consider
\begin{align}
    \rho=\begin{pmatrix}
1 & 0\\ 0 & 0
    \end{pmatrix}.
\end{align}
Its square root gives  $W=\sqrt{\rho}=\rho$. Thus, the polar decomposition of $W$ is $W=\sqrt{\rho}U$ with
\begin{align}
    U=\begin{pmatrix}
1 & 0\\ 0 & \me^{\mi\chi}
    \end{pmatrix}.
\end{align}
Here $\chi$ is an arbitrary real number. Physically, the purification $W$ no longer has an unique phase factor, in contrast to the full-rank case. The arbitrariness of the polar decomposition of the amplitude thus hinders the construction of a bundle like the Uhlmann bundle. This is a severe obstacle when generalizing the previous approach to $\mathcal{D}^N_{k<N}$.

For density matrices $\rho_1$ and $\rho_2$ each of rank $k<N$ in $\mathcal{D}^{N}_{k}$, one may generalize the Bures distance shown in Eq. (\ref{gyy3}) to $\dif^2_\text{B}(\rho_1,\rho_2)=\inf\text{Tr}[(W_1-W_2)(W_1-W_2)^\dag]$. However, the condition to saturate the infimum will be different from the parallel-transport condition shown in Eq. (\ref{pcm}). This is because $W_{1}$ and $W_2$ are not full-rank, so their zero-eignevalues may invalidate the condition. Alternatively, one may take Eq.~(\ref{db3}) as another definition of a local distance, but this distance is no longer the minimum of Eq. (\ref{gyy3}). This is in contrast to the full-rank case, where the two expressions agree with each other. Consequently, the precise local geometry $\mathcal{D}^N_k$ can be different from that of $\mathcal{D}^N_N$. Moreover, the construction of the Uhlmann bundle no longer applies if the rank is not full since the polar decomposition of the amplitude is no longer unique, so a geometric foundation for $\mathcal{D}^N_{k<N}$ is lacking. Therefore, the discovery of a gauge-invariant QGT for $\mathcal{D}^N_{k<N}$ will be a challenge. Nevertheless, this scenario underscores the need for more research on the local geometry of mixed states.

\section{Conclusion}\label{sec:Con}
Through an analogue of the QGT of pure states and purification of density matrices, we explicitly construct the QGT of mixed states applicable to thermal equilibrium based on the Uhlmann fibration. The Pythagorean-like equations of pure and mixed states relate the distances in the phase space and its embedded space and reveal the corresponding parallel-transport conditions. The gauge-invariant modification of the mixed-state QGT leads to the Bures metric in the real part and the Uhlmann form in the imaginary part. While the Bures metric is the natural result within the framework, the Uhlmann form vanishes for common systems.
The Bures metric reduces to the Fubini-Study metric of pure states as temperature approaches zero, but the correspondence only applies to the real part of the QGT.
The QGT of mixed states not only characterizes the local geometry of quantum statistical systems but also allows us to explore geometric effects beyond the ground state in rapidly developed quantum devices and simulators.

\section{Acknowledgments}
We thank Prof. Guang-Yu Guo for bringing our attention to the QGT. H.G. was supported by the National Natural Science
Foundation of China (Grant No. 12074064) and the Innovation Program for Quantum
Science and Technology (Grant No. 2021ZD0301904). C.C.C. was supported
by the National Science Foundation under Grant No.
PHY-2310656 and thanks the NCTS, Taiwan for its hospitality during his visit. X. Y. H. was supported by the Jiangsu Funding Program for Excellent Postdoctoral Talent (Grant No. 2023ZB611).

\appendix

\section{Gauge invariant inner product}\label{app:inner}
The metric is extracted from the distance defined by an inner product. Therefore, we first build a suitable scalar product on the phase space $P(\mathcal{H})$. Since the metric tensor is a bilinear map on the tangent vectors, we define the inner product as
\begin{align}\label{Hip}
\langle\cdot,\cdot\rangle: T_{|\psi\rangle}P(\mathcal{H})\times T_{|\psi\rangle}P(\mathcal{H})\rightarrow \mathbb{C},
\end{align}
where $T_{|\psi\rangle}P(\mathcal{H})$ is the tangent space at the point $|\psi\rangle\in P(\mathcal{H})$.
Since $|\psi\rangle$ actually represents an equivalent class with respect to the equivalence relation $|\psi\rangle=\me^{\mi\chi}|\phi\rangle\sim |\phi\rangle$, the scalar product $\langle\cdot,\cdot\rangle$ must be gauge invariant. There is a natural way to accomplish this. Since $T_{|\psi\rangle}P(\mathcal{H})\cong P(\mathcal{H})$, any point $X\in T_{|\psi\rangle}P(\mathcal{H})$ can be equivalently mapped to a state-vector $|X\rangle\in P(\mathcal{H})$:
If $X$ is a tangent vector of a curve parameterized by $t$, then $X=\frac{\dif}{\dif t}$ and $\label{Xt}|X(t)\rangle\equiv X|\psi(t)\rangle=\frac{\dif}{\dif t}|\psi(t)\rangle$, implying $|\psi(t+\dif t)\rangle=|\psi(t)\rangle+\dif t|X(t)\rangle$ up to first order of $\dif t$. Thus, the inner product between $X, Y\in T_{|\psi\rangle}P(\mathcal{H})$ is equivalently expressed as $\langle X, Y\rangle=\langle X|Y\rangle$, where $\langle\cdot|\cdot\rangle$ is the inner product between pure states. This provides a gauge-invariant inner product on $P(\mathcal{H})$.

By choosing $|\tilde{\psi}\rangle\in \Pi^{-1}(|\psi\rangle)$, $X$ can be obtained by projecting a vector $\tilde{X}\in T_{|\tilde{\psi}\rangle}\mathcal{H}$ via $X=\Pi_*(\tilde{X})$, where $\Pi_*$ is the push-forward induced by $\Pi$. Similarly, $\tilde{X}$ can also be represented by $|\tilde{X}\rangle$. Furthermore, we decompose $\tilde{X}$ into the parallel and perpendicular parts with respect to $|\tilde{\psi}\rangle$ as
\begin{align}\label{Xt3}
|\tilde{X}\rangle=\lambda|\tilde{\psi}\rangle+|\tilde{X}^\perp\rangle,
\end{align}
where $\lambda=\frac{\langle\tilde{\psi}|\tilde{X}\rangle}{\langle\tilde{\psi}|\tilde{\psi}\rangle}$, and $|\tilde{X}^\perp\rangle$ satisfies $\langle\tilde{\psi}|\tilde{X}^\perp\rangle=0$. 
Under the gauge transformation $|\tilde{\psi}'\rangle=\me^{\mi\chi}|\tilde{\psi}\rangle$, the perpendicular component of $|\tilde{X}\rangle$ is always invariant:
\begin{align}
|\tilde{X}'^\perp\rangle&=|\tilde{X}\rangle-\frac{\langle \tilde{\psi}'|\tilde{X}\rangle}{\langle\tilde{\psi}'|\tilde{\psi}'\rangle}|\tilde{\psi}'\rangle=|\tilde{X}^\perp\rangle.
\end{align}
Thus, for any pair of vectors $X_1$, $X_2\in T_{|\psi\rangle}P(\mathcal{H})$, a gauge invariant scalar product between them can be defined as
\begin{align}\label{Hsp}
\langle X_1|X_2\rangle &:=\frac{\langle\tilde{ X}^\perp_1|\tilde{X}^\perp_2\rangle}{\langle\tilde{\psi}|\tilde{\psi}\rangle} \\ \notag
&=\frac{\langle\tilde{ X}_1|\tilde{X}_2\rangle\langle\tilde{\psi}|\tilde{\psi}\rangle-\langle\tilde{X}_1|\tilde{\psi}\rangle\langle\tilde{\psi}|\tilde{X}_2\rangle}{\langle\tilde{\psi}|\tilde{\psi}\rangle^2}.
\end{align}
Here Eq.~(\ref{Xt3}) has been applied.
If $|\tilde{\psi}\rangle$ is restricted to $S(\mathcal{H})$, then
\begin{align}\label{X12a}
\langle X_1|X_2\rangle=\langle\tilde{ X}_1|\tilde{X}_2\rangle-\langle\tilde{X}_1|\tilde{\psi}\rangle\langle\tilde{\psi}|\tilde{X}_2\rangle.
\end{align}
This is the desired gauge-invariant inner product for Eq.~(\ref{Hip}), which further induces a norm $||X||:=\langle X,X\rangle$.

\section{K$\ddot{\text{a}}$hler metric}\label{appKahler}
The K$\ddot{\text{a}}$hler manifold is naturally equipped with a K$\ddot{\text{a}}$hler metric. However, the previously introduced $g_{ij}$ in Sec.~\ref{SecII} is not of this type because we have initially chosen a set of real-valued parameters rather than instilling a complex structure. By employing a set of local complex coordinates, the K$\ddot{\text{a}}$hler metric of $CP^{N-1}$ can be derived. This has been achieved \cite{Nomizu_Book,Nakahara} in an abstract manner, but here we present a more physical demonstration starting from Eq.~(\ref{FSm1}) and
relaxing the restriction  $|\tilde{\psi}\rangle \in S(\mathcal{H})$. Therefore, $\langle\tilde{\psi}|\tilde{\psi}\rangle\neq 1$ in general, and Eq.~(\ref{FSm1}) now takes the form
\begin{align}\label{FSm2}
\langle \partial_i\psi|\partial_j\psi\rangle=\frac{\langle \partial_i\tilde{\psi}|\partial_j\tilde{\psi}\rangle\langle\tilde{\psi}|\tilde{\psi}\rangle-\langle \partial_i\tilde{\psi}|\tilde{\psi}\rangle\langle\tilde{\psi}|\partial_j\tilde{\psi}\rangle}{\langle\tilde{\psi}|\tilde{\psi}\rangle^2}.
\end{align}
We choose an orthonormal basis of $\mathbb{C}^N$, under which the coordinates of $|\tilde{\psi}\rangle \in \mathbb{C}^N$ can be represented by a generic complex vector $(z^0,z^1,\cdots,z^{N-1})^T$, such that
\begin{align}
r^2=\langle\tilde{\psi}|\tilde{\psi}\rangle=\delta_{\alpha\beta}\bar{z}^\alpha z^\beta =\bar{z}_\alpha z^\alpha.
\end{align}
Here we use the Greek alphabets $\alpha,\beta=0,1,2,\cdots,N-1$ to label the coordinate components. The elements of the flat metric $\delta^{\alpha\beta}$ or $\delta_{\alpha\beta}$ are used to raise or lower the indices.
Since $CP^{N-1}=\mathbb{C}^{N}/\mathbb{C}^*$, every point in $CP^{N-1}$ represents a class in $\mathbb{C}^{N}$ with respect to any equivalence (gauge) transformation $c\in \mathbb{C}^*$. Explicitly, $|\tilde{\psi}\rangle\sim|\tilde{\psi}'\rangle$ if $|\tilde{\psi}\rangle= c|\tilde{\psi}'\rangle$ or $z^\alpha=cz^{\prime \alpha}$. We take $z^0\neq 0$ without loss of generality, and the coordinates of $|\psi\rangle\in CP^{N-1}$ are thereby obtained via the projection
\begin{align}\label{wi}
w^i=\frac{z^i}{z^0},\quad i=1,2,\cdots,N-1.
\end{align}
Here $w^i$s are called the homogeneous coordinates, whose indices are labelled by Latin alphabets.

Since $\partial_i\equiv \frac{\partial}{\partial w_i}=z^0\frac{\partial}{\partial z^i}$ and $|\tilde{\psi}\rangle=(z^0,z^1,\cdots,z^{N-1})^T$, $|\partial_i\tilde{\psi}\rangle=z^0(0,\cdots,1,\cdots,0)^T$, where only the $i$th component is nonzero. Substituting these into Eq.~(\ref{FSm2}), we get
\begin{align}\label{FSm3}
\langle \partial_i\psi|\partial_j\psi\rangle&=\frac{\delta_{ij}\bar{z}_0z^0}{\bar{z}_\alpha z^\alpha}-\frac{\bar{z}_0z^0z_i\bar{z}_j}{(\bar{z}_\alpha z^\alpha)^2}\notag\\&=\frac{(1+\bar{w}_kw^k)\delta_{ij}-w_i\bar{w}_j}{(1+\bar{w}_kw^k)^2},
\end{align}
where we have applied $\bar{z}_\gamma z^\gamma=(1+\bar{w}_kw^k)\bar{z}_0z^0$ in the second step.
Thus, the local distance in $CP^{N-1}$ is
\begin{align}\label{CPNs0}
\dif s^2=\langle \dif\psi|\dif\psi\rangle=\frac{(1+\bar{w}_kw^k)\delta_{ij}-w_i \bar{w}_j}{(1+\bar{w}_kw^k)^2}\dif \bar{w}^i\dif w^j.
\end{align}
Expressing $\dif s^2=2g_{i\bar{j}}\dif \omega^i\dif \bar{\omega}^j$, we get
\begin{align}\label{FSm4}
g_{i\bar{j}}&=\frac{1}{2}\frac{(1+\bar{w}_kw^k)\delta_{ij}-\bar{w}_iw_j}{(1+\bar{w}_kw^k)^2},\notag\\
g_{ij}&=g_{\bar{i}\bar{j}}=0.
\end{align}
Since $\dif s^2$ is real, it can also be expressed as
\begin{align}\label{tmp2}
\dif s^2=g_{i\bar{j}}\dif \omega^i\dif \bar{\omega}^j+\overline{g_{i\bar{j}}}\dif \bar{\omega}^i\dif w^j,
\end{align}
which implies $g_{\bar{i}j}=\overline{g_{i\bar{j}}}$, so $g_{\bar{i}j}$ is Hermitian. In differential geometry, $g_{\bar{i}j}$ is the component form of the K$\ddot{\text{a}}$hler metric on $CP^{N-1}$ \cite{Nakahara}.
Introducing the K$\ddot{\text{a}}$hler potential $K=\ln\sqrt{1+\bar{w}_kw^k},$ it can be further expressed as $g_{i\bar{j}}=\frac{\partial^2 K}{\partial w^i\partial\bar{w}^j}.$


It has been pointed out that the real and imaginary parts of QGT are the Fubini-Study metric and Berry curvature, respectively \cite{QGTCMP80,QGT10}. Here we show how this arises from the K$\ddot{\text{a}}$hler geometry. Different from our previous discussions, complex coordinates must be included in order to yield the correct result. In terms of differential forms, the local distance $\dif s^2=2g_{i\bar{j}}\dif \omega^i\dif \bar{\omega}^j$ on $CP^{N-1}$ suggests a tensor field $\mathcal{G}=2g_{i\bar{j}}\dif \omega^i\otimes\dif \bar{\omega}^j$, which is non-Hermitian and plays the role of QGT with complex coordinates. On the other hand, the Hermitian form (\ref{tmp2}) of $\dif s^2$ introduces
the K$\ddot{\text{a}}$hler metric
\begin{align}\label{tmp3}
g=g_{i\bar{j}}\dif \omega^i\otimes\dif \bar{\omega}^j+g_{\bar{i}j}\dif \bar{\omega}^i\otimes\dif w^j,
\end{align}
which is the real part of $\mathcal{G}$ and gives the Fubini-Study metric in the form of complex coordinates. Since $CP^{N-1}$ is a complex manifold, it admits an almost complex structure $J$ satisfying $J\frac{\partial}{\partial w^i}=\mi \frac{\partial}{\partial w^i}$ and $J\frac{\partial}{\partial \bar{w}^i}=-\mi \frac{\partial}{\partial \bar{w}^i}$ \cite{Nakahara}. With this, one can define the K$\ddot{\text{a}}$hler form whose action on $X,Y \in TP(\mathcal{H})$ is
$\Omega(X,Y):=g(JX,Y)$. A straightforward calculation shows
\begin{align}\label{F}
\Omega&=\mi g_{i\bar{j}}\dif \omega^i\otimes\dif \bar{\omega}^j-\mi g_{\bar{i}j}\dif \bar{\omega}^i\otimes\dif w^j\notag\\&=\mi g_{i\bar{j}}\dif \omega^i\wedge\dif \bar{\omega}^j,
\end{align}
which exactly agrees with the (negative) imaginary part of $\mathcal{G}$. Moreover, $\Omega$ is proportional to the Berry curvature
\begin{align}\label{FAA}
F=\dif A=\frac{\bar{w}_i w_j-(1+\bar{w}_kw^k)\delta_{ij}}{(1+\bar{w}_kw^k)^2}\dif w^i\wedge \dif \bar{w}^j.
\end{align}
Here the Berry connection for an unnormalized state $|\tilde{\psi}\rangle$ is
\begin{align}\label{AAp2}
A&=\frac{\mi\text{Im}\langle\tilde{\psi}|\dif|\tilde{\psi}\rangle}{\langle\tilde{\psi}|\tilde{\psi}\rangle}=\frac{\mi\text{Im}(\bar{w}_i\dif w^i)}{1+\bar{w}_kw^k}
=\frac{1}{2}\frac{\bar{w}_i\dif w^i-w^i\dif\bar{w}_i}{1+\bar{w}_kw^k}.
\end{align}
From Eqs.~(\ref{FSm4}) and (\ref{F}), we indeed have $\Omega=-\frac{\mi}{2}F$. Therefore, the QGT of pure states has a profound geometrical origin.

We present a simple example of a two-level system with $N=2$ that corresponds to the Hopf fibration: $S^3/$U(1)$=CP^1\cong S^2$. A quantum state in $S^3$ is expressed by $(z^0,z^1)^T$ with $|z^0|^2+|z^1|^2=1$. Hence, it can be parameterized by three real parameters. Let $z^0=\me^{\mi(\chi-\frac{\phi}{2})}\cos\frac{\theta}{2}$ and $z^1=\me^{\mi(\chi+\frac{\phi}{2})}\cos\frac{\theta}{2}$. There is only one homogeneous coordinate $w=\frac{z^1}{z^0}=\me^{\mi\phi}\tan\frac{\theta}{2}.$ The K$\ddot{\text{a}}$hler potential is $K=\frac{1}{2}\ln(1+\bar{w}w)$, from which we obtain the component of the Fubini-Study metric:
\begin{align}
g_{w\bar{w}}
=-\frac{1}{2}\frac{\bar{w}w}{(1+\bar{w}w)^2}+\frac{1}{2}\frac{1}{1+\bar{w}w}=\frac{1}{2}\cos^4\frac{\theta}{2}.
\end{align}
The Fubini-Study distance is thus given by
\begin{align}\label{ds2CP1}
\dif s^2=2g_{w\bar{w}}\dif w\dif\bar{w}=\frac{1}{4}\left(\dif\theta^2+\sin^2\theta\dif\phi^2\right),
\end{align}
which is exactly the local distance on the two-dimensional sphere.
The QGT is
\begin{align}\label{QGTps}
\mathcal{G}&=2g_{w\bar{w}}\dif w\otimes\dif \bar{w}\notag\\
&=\cos^4\frac{\theta}{2}\Big[\tan^2\frac{\theta}{2}\dif\phi\otimes\dif\phi+\frac{1}{4}\sec^4\frac{\theta}{2}\dif\theta\otimes\dif\theta\notag\\
&-\frac{\mi}{2}\sec^2\frac{\theta}{2}\tan\frac{\theta}{2}(\dif\theta\otimes\dif\phi-\dif\phi\otimes \dif\theta)\Big]\notag\\
&=\frac{1}{4}\left(\dif\theta\otimes\dif\theta+\sin^2\theta\dif\phi\otimes\dif\phi\right)-\frac{\mi}{4}\sin\theta\dif\theta\wedge\dif\phi.
\end{align}
The real part is the Fubini-Study metric while the (negative) imaginary part is proportional to the Berry curvature of the two-level system  \cite{HBY_book}
\begin{align}
F&=\frac{\mi}{2}\sin\theta\dif\theta\wedge\dif\phi.
\end{align}

\section{Properties of the space of mixed states}\label{app0}
We have denoted by $\mathcal{P}$ the space of all density matrices. For $\rho_1$, $\rho_2\in\mathcal{P}$ and a pair of arbitrary complex numbers $\lambda$ and $\mu$, it is possible to find counterexamples to $\lambda\rho_1+\mu\rho_2\in \mathcal{P}$. On the other hand, as a complex matrix, $\rho$ is also equivalent to an operator acting on the $N$-tuple vectors in $\mathcal{H}=\mathbb{C}^N$. Consequently, $\rho $ belongs to $End(\mathcal{H})$, which is the space of operators on $\mathcal{H}$, and is also known as the algebra of $N\times N$ complex matrices. This means $\mathcal{P}\subset End(\mathcal{H})$. Furthermore, the requirement that $\text{Tr}(\lambda\rho_1+\mu\rho_2)=1$ leads to $\lambda+\mu=1$, which further implies $\mathcal{P}$ is a convex subset of $\mathcal{P}\subset End(\mathcal{H})$  \cite{Bengtsson_book}.

To determine the dimension of $\mathcal{P}$, we note that a complex $N\times N$ matrix $\rho$ contains $N^2$ complex parameters or $2N^2$ real parameters. In addition, the condition $\text{Tr}\rho=1$ provides one restriction, and the Hermiticity requires that (1) the diagonal elements are real, adding $N$ restrictions, and (2) the off-diagonal elements satisfy $\rho_{ij}=\rho^*_{ji}$, imposing $2\times\frac{N(N-1)}{2}$ more restrictions. Thus, $\dim \mathcal{P}=2N^2-N(N-1)-N-1=N^2-1$. This implies that any $\rho$ can be mapped to a real $N\times N$ matrix $\tilde{\rho}$ such that $\text{Tr}\tilde{\rho}=1$ and rank($\rho$)=rank($\tilde{\rho}$). We now use this mapping to determine the dimension of $\mathcal{D}^N_k$. If rank$(\tilde{\rho})=k$, it can be factorized as $\tilde{\rho}=AB$, where $A$ is a $N\times k$ matrix and $B$ is a $k\times N$ matrix. This factorization is not unique since for any $k\times k$ invertible matrix $R$, $\tilde{\rho}=(AR)(R^{-1}B)$. This introduces $k^2$ redundant degrees of freedom. The factorization itself has $2Nk-1$ parameters, so the total number of independent parameters is $2Nk-k^2-1=N^2-(N-k)^2-1=\dim \mathcal{D}^N_k$.

\section{Infimum of Eq.~(\ref{gyy2})}\label{appa0}
The infimum of Eq.~(\ref{gyy2}) leads to the definition of the Bures distance (\ref{gyy3}), which is evaluated as
\begin{align}\label{db2}
\dif^2_\textrm{B}(\rho(t),\rho(0))=2-2\sup\text{Re}\textrm{Tr}\left[W^\dagger(0)W(t)\right].
\end{align}
Let $A=W^\dag(0)W(t)$, which is also full rank and thereby has a unique decomposition $A=|A|U_A$ with $|A|=\sqrt{AA^\dagger}$. Applying the Cauchy-Schwartz inequality $|\textrm{Tr}(A^\dagger B)|^2\le \textrm{Tr}(A^\dagger A)\textrm{Tr}(B^\dagger B)$, we get
\begin{align}\label{ieq}
\textrm{Re}\big[\textrm{Tr}(A)\big]&\leq |\textrm{Tr}(A)|=|\textrm{Tr}(\sqrt{|A|}\sqrt{|A|}U_A)|\notag\\
&\leq \sqrt{\textrm{Tr}|A|\textrm{Tr}\big(U^\dagger_A|A|U_A\big)}=\textrm{Tr}|A|,
\end{align}
where the inequality is saturated whenever $\sqrt{|A|}=\sqrt{|A|}U_A$. This leads to
\begin{align}A=|A|=\sqrt{AA^\dag}&\Longrightarrow A^2=AA^\dag\notag\\&\Longrightarrow A=A^\dag=|A|>0,\end{align}
i.e.,
\begin{align}\label{pcmb}
W^\dag(0)W(t)=W^\dag(t)W(0)>0.
\end{align}
The Bures distance is
\begin{align}\label{db2b}
\dif^2_\textrm{B}(\rho(t),\rho(0))&=2-2\textrm{Tr}|W^\dag(0)W(t)|\notag\\
&=2-2\textrm{Tr}\sqrt{W^\dag(0)W(t)W^\dag(t)W(0)}\notag\\
&=2-2\textrm{Tr}\sqrt{\sqrt{\rho}\rho(t)\sqrt{\rho}},
\end{align}
where we have applied the fact $W(0)=W^\dag(0)=\sqrt{\rho}$.

Substitute $W(t)=W(0)+tV$ and $W(0)=\sqrt{\rho}$ into the condition (\ref{pcmb}), we further get
  \begin{align}\label{pcmb2}
V^\dag\sqrt{\rho}=\sqrt{\rho}V,
\end{align}
which means $V\px \sqrt{\rho}$. Both $V$ and $\rho$ are full-ranked, then Eq.~(\ref{pcmb2}) implies $V^\dag=\sqrt{\rho}V\sqrt{\rho}^{-1}$. Plugging it into $\sqrt{\rho}V^\dag+V\sqrt{\rho}=Y_\rho$, we have
  \begin{align}\label{pcmb3}
\rho V+V\rho=Y_\rho\sqrt{\rho}.
\end{align}
Multiply both sides with the completeness relation of the eigenstates of $\rho$: $\sum_i|i\rangle\langle i|=1$, we obtain
  \begin{align}\label{pcmb4}
\sum_{ij}(\lambda_i+\lambda_j)\langle i|V|j\rangle|i\rangle\langle j|=\sum_{ij}\sqrt{\lambda_j}\langle i|Y_\rho|j\rangle|i\rangle\langle j|.
\end{align}
Due to the linear independence of $\{|i\rangle\langle j|\}$, we have
  \begin{align}\label{pcmb5}
\langle i|V|j\rangle=\frac{\sqrt{\lambda_j}}{\lambda_i+\lambda_j}\langle i|Y_\rho|j\rangle.
\end{align}
Thus, when
  \begin{align}\label{pcmb6}
V=\sum_{ij}\frac{\sqrt{\lambda_j}}{\lambda_i+\lambda_j}\langle i|Y_\rho|j\rangle|i\rangle\langle j|,
\end{align}
$ g(Y_\rho,Y_\rho)$ reaches its infimum, which defines the Bures distance.

\section{Details of the Bures distance}\label{appa}
Following Ref.~\cite{HubnerPLA92}, we set $A(t)=\sqrt{\sqrt{\rho}(\rho+t\mathrm{d}\rho)\sqrt{\rho}}$, then $A(0)=\rho.$
The squared Bures distance can be expanded  up to second order in $t$ as $\dif^2_\textrm{B}(\rho,\rho+t\mathrm{d}\rho)=t^2g^\text{B}_{\mu\nu}(\rho)\dif R^\mu\dif R^\nu$. Thus Eq.~(\ref{db3}) implies
\begin{align}\label{A2}
g^\text{B}_{\mu\nu}(\rho)\dif R^\mu\dif R^\nu=\frac{1}{2}\frac{\mathrm{d}^2}{\mathrm{d}t}\dif^2_\textrm{B}(\rho,\rho+t\mathrm{d}\rho)\big|_{t=0}=-\textrm{Tr}\ddot{A}(t)\big|_{t=0}.
\end{align}
Since
\begin{align}\label{A3}
A(t)A(t)=\sqrt{\rho}(\rho+t\mathrm{d}\rho)\sqrt{\rho},
\end{align}
Differentiating both sides twice with respect to $t$, we get
\begin{align}\label{A4}
&\dot{A}(0)A(0)+A(0)\dot{A}(0)=\sqrt{\rho}\mathrm{d}\rho\sqrt{\rho},\notag\\
&\ddot{A}(0)A(0)+2\dot{A}(0)\dot{A}(0)+A(0)\ddot{A}(0)=0.
\end{align}
Multiplying $A^{-1}(0)=\rho^{-1}$ from the left on both sides of the second equation and taking trace, we have
\begin{align}\label{A6}
\textrm{Tr}\big(\rho^{-1}\ddot{A}(0)\rho\big)+2\textrm{Tr}\big[\rho^{-1}\big(\dot{A}(0)\big)^2\big]+\textrm{Tr}\ddot{A}(0)=0,
\end{align}
which leads to
\begin{align}\label{A7}
\textrm{Tr}\ddot{A}(0)=-\textrm{Tr}\big[\rho^{-1}\big(\dot{A}(0)\big)^2\big].
\end{align}
Using $\rho|i\rangle=\lambda_i|i\rangle$ and $\sqrt{\rho}|i\rangle=\sqrt{\lambda_i}|i\rangle$, the first equation of (\ref{A4}) implies
\begin{align}\label{A9}
(\lambda_i+\lambda_j)\langle i|\dot{A}(0)|j\rangle=\sqrt{\lambda_i\lambda_j}\langle i|\mathrm{d}\rho|j\rangle.
\end{align}
Therefore by applying Eqs.~(\ref{A2}) and (\ref{A7}), we have
\begin{align}\label{A10}
\dif^2_\textrm{B}(\rho,\rho+t\mathrm{d}\rho)&=-\textrm{Tr}\ddot{A}(0)t^2\notag\\
&=\sum_{ij}\frac{1}{\lambda_i}\langle i|\dot{A}(0)|j\rangle\langle j|\dot{A}(0)|i\rangle t^2\notag\\
&=\sum_{ij}\frac{1}{\lambda_i}\frac{\lambda_i\lambda_j}{(\lambda_i+\lambda_j)^2}|\langle i|\mathrm{d}\rho|j\rangle|^2 t^2,
\end{align}
Now interchange the indices $i$ and $j$ and setting $t=1$, we have
\begin{align}\label{A11b}
\dif^2_\textrm{B}(\rho,\rho+\mathrm{d}\rho)
&=\frac{1}{2}\sum_{ij}\frac{|\langle i|\mathrm{d}\rho|j\rangle|^2}{\lambda_i+\lambda_j}.
\end{align}

The Bures distance has another equivalent expression that is useful to our discussions.
Using $\sqrt{\rho}|i\rangle=\sqrt{\lambda_i}|i\rangle$ and $\mathrm{d}\rho=\mathrm{d}(\sqrt{\rho}\sqrt{\rho})=\mathrm{d}\sqrt{\rho}\sqrt{\rho}+\sqrt{\rho}\mathrm{d}\sqrt{\rho}$, the matrix element of $\dif\rho$ becomes
\begin{align}
\langle i|\mathrm{d}\rho|j\rangle=(\sqrt{\lambda_i}+\sqrt{\lambda_j})\langle i|\mathrm{d}\sqrt{\rho}|j\rangle.
\end{align}
Substituting it into Eq.~(\ref{A11b}), we get
 \begin{align}\label{Bmb2}
\dif^2_\textrm{B}(\rho,\rho+\mathrm{d}\rho)=\frac{1}{2}\sum_{ij}\frac{(\sqrt{\lambda_i}+\sqrt{\lambda_j})^2}{\lambda_i+\lambda_j}|\langle i|\dif\sqrt{\rho}|j\rangle|^2.
\end{align}

In the two-dimensional case, the Bures distance takes a simpler expression. Let $\rho=\frac{1}{2}+\mathbf{a}\cdot\boldsymbol{\sigma}$, then its eigenvalues are $\lambda_{1,2}=\frac{1}{2}\pm |\mathbf{a}|$ and the projective operators for each eigenvector are $|1\rangle\langle1|=\frac{1}{2}\left(1+\hat{\mathbf{a}}\cdot\boldsymbol{\sigma}\right)$, $|2\rangle\langle2|=\frac{1}{2}\left(1-\hat{\mathbf{a}}\cdot\boldsymbol{\sigma}\right)$ with $\hat{\mathbf{a}}=\frac{\mathbf{a}}{|\mathbf{a}|}$. Using $\dif\rho=\dif\mathbf{a}\cdot\boldsymbol{\sigma}$ and $\lambda_1+\lambda_2=1$, the Bures distance becomes
\begin{widetext}
\begin{align}\label{A11c}
\dif^2_\textrm{B}(\rho,\rho+\mathrm{d}\rho)&=\frac{1}{4}\sum_{i}\frac{|\langle i|\mathrm{d}\rho|i\rangle|^2}{\lambda_i}+\frac{1}{2}\sum_{i\neq j}|\langle i|\mathrm{d}\rho|j\rangle|^2\notag\\
&=\langle1|\dif\mathbf{a}\cdot\boldsymbol{\sigma}\left(\frac{1+\hat{\mathbf{a}}\cdot\boldsymbol{\sigma}}{8\lambda_1}+\frac{1-\hat{\mathbf{a}}\cdot\boldsymbol{\sigma}}{4}\right)\dif\mathbf{a}\cdot\boldsymbol{\sigma}|1\rangle+
\langle2|\dif\mathbf{a}\cdot\boldsymbol{\sigma}\left(\frac{1-\hat{\mathbf{a}}\cdot\boldsymbol{\sigma}}{8\lambda_2}+\frac{1+\hat{\mathbf{a}}\cdot\boldsymbol{\sigma}}{4}\right)\dif\mathbf{a}\cdot\boldsymbol{\sigma}|2\rangle\notag\\
&=\left(\frac{1}{8\lambda_1\lambda_2}+\frac{1}{2}\right)\dif\mathbf{a}\cdot\dif\mathbf{a}-\sum_{i=1,2}(-1)^i\frac{1-2\lambda_i}{8\lambda_i}\langle i|\dif\mathbf{a}\cdot\boldsymbol{\sigma}\hat{\mathbf{a}}\cdot\boldsymbol{\sigma}\dif\mathbf{a}\cdot\boldsymbol{\sigma}|i\rangle.
\end{align}
\end{widetext}
Next, applying $\langle 1,2|\hat{\mathbf{a}}\cdot\boldsymbol{\sigma}|1,2\rangle=\pm 1$, $1-2\lambda_{1,2}=\mp2|\mathbf{a}|$, and
\begin{align}
\dif\mathbf{a}\cdot\boldsymbol{\sigma}\hat{\mathbf{a}}\cdot\boldsymbol{\sigma}\dif\mathbf{a}\cdot\boldsymbol{\sigma}&=\left[\dif \mathbf{a}\cdot\hat{\mathbf{a}}+\mi(\dif \mathbf{a}\times\hat{\mathbf{a}})\cdot\boldsymbol{\sigma}\right]\dif\mathbf{a}\cdot\boldsymbol{\sigma}\notag\\&=2(\dif \mathbf{a}\cdot\hat{\mathbf{a}})\dif \mathbf{a}\cdot\boldsymbol{\sigma}-(\dif \mathbf{a}\cdot\dif \mathbf{a})\hat{\mathbf{a}}\cdot\boldsymbol{\sigma},\notag
\end{align}
we further get
\begin{align}\label{A11d}
\dif^2_\textrm{B}(\rho,\rho+\mathrm{d}\rho)=&\left(1+\frac{1}{8\lambda_1\lambda_2}-\frac{1}{8\lambda_1}-\frac{1}{8\lambda_2}\right)\dif\mathbf{a}\cdot\dif\mathbf{a}\notag\\
-&\dif\mathbf{a}\cdot\mathbf{a}\left(\frac{\dif \mathbf{a}\cdot\langle 1|\boldsymbol{\sigma}|1\rangle}{2\lambda_1}+\frac{\dif \mathbf{a}\cdot\langle 2|\boldsymbol{\sigma}|2\rangle}{2\lambda_2}\right)\notag\\
=&\dif\mathbf{a}\cdot\dif\mathbf{a}-\frac{(\dif\mathbf{a}\cdot\mathbf{a})^2}{2|\mathbf{a}|}\left(\frac{1}{\lambda_1}-\frac{1}{\lambda_2}\right)\notag\\
=&\dif\mathbf{a}\cdot\dif\mathbf{a}+\frac{(\dif\mathbf{a}\cdot\mathbf{a})^2}{\det\rho},
\end{align}
where $\langle 1|\boldsymbol{\sigma}|1\rangle=-\langle 2|\boldsymbol{\sigma}|2\rangle=\hat{\mathbf{a}}$ has been applied. Moreover, noting
\begin{align}\label{t2}
\textrm{Tr}(\mathrm{d}\rho)^2&=\textrm{Tr}( \mathrm{d}\mathbf{a}\cdot\boldsymbol{\sigma})^2=2\mathrm{d}\mathbf{a}\cdot \mathrm{d}\mathbf{a},
\end{align}
and
\begin{align}\label{t3}
\dif\sqrt{\det\rho}=\dif\sqrt{\frac{1}{4}-\mathbf{a}^2}=-\frac{\dif\mathbf{a}\cdot\mathbf{a}}{\sqrt{\det\rho}},
\end{align}
it is found that in the two-dimensional case,
\begin{align}\label{A11e}
\dif^2_\textrm{B}(\rho,\rho+\mathrm{d}\rho)=\frac{1}{2}\textrm{Tr}(\mathrm{d}\rho)^2+\left(\dif\sqrt{\det\rho}\right)^2.
\end{align}

\section{Modification of the real and imaginary parts of $\gamma_{\mu\nu}$}\label{appc}
Under the gauge transformation $W'=W\mathcal{U}$, 
\begin{align}
\gamma'_{\mu\nu}=&\gamma_{\mu\nu}+\frac{1}{2}\text{Tr}\left[(W^\dag\partial_\nu W-\partial_\nu W^\dag W)\mathcal{U}\partial_\mu \mathcal{U}^\dag\right]\notag\\
+&\frac{1}{2}\text{Tr}\left[(W^\dag\partial_\mu W-\partial_\mu W^\dag W)\mathcal{U}\partial_\nu \mathcal{U}^\dag\right]\notag\\
+&\frac{1}{2}\text{Tr}\left[W^\dag W(\partial_\mu \mathcal{U}\partial_\nu \mathcal{U}^\dag+\partial_\nu \mathcal{U}\partial_\mu \mathcal{U}^\dag)\right],\label{g1}\\
\sigma'_{\mu\nu}=&\sigma_{\mu\nu}+\frac{\mi}{2}\text{Tr}\left[\partial_\mu(W^\dag W)\mathcal{U}\partial_\nu \mathcal{U}^\dag-\partial_\nu(W^\dag W)\mathcal{U}\partial_\mu \mathcal{U}^\dag\right]\notag\\
+&\frac{\mi}{2}\text{Tr}\left[W^\dag W(\partial_\mu \mathcal{U}\partial_\nu \mathcal{U}^\dag-\partial_\nu \mathcal{U}\partial_\mu \mathcal{U}^\dag)\right],\label{sigma1}\\
\omega'_\mu=&\mathcal{U}^\dag\omega_\mu \mathcal{U}+\mathcal{U}^\dag \partial_\mu \mathcal{U}.\label{ot2}
\end{align}
Here $\omega_\mu$ is the component form of $\omega$ when restricted on $\mathcal{D}^N_N$, which satisfies \cite{GPbook,OurDPUP}
\begin{align}\label{wdwo1}
W^\dagger\partial_\mu W-\partial_\mu W^\dagger W=W^\dagger W\omega_\mu+\omega_\mu W^\dagger W.
\end{align}
Using this, the transformation (\ref{g1}) becomes
\begin{align}\label{g2}
\gamma'_{\mu\nu}=&\gamma_{\mu\nu}+\frac{1}{2}\text{Tr}\left[(W^\dagger W\omega_\mu+\omega_\mu W^\dagger W)\mathcal{U}\partial_\nu \mathcal{U}^\dag\right]\notag\\
+&\frac{1}{2}\text{Tr}\left[(W^\dagger W\omega_\nu+\omega_\nu W^\dagger W)\mathcal{U}\partial_\mu \mathcal{U}^\dag\right]\notag\\
+&\frac{1}{2}\text{Tr}\left[W^\dag W(\partial_\mu \mathcal{U}\partial_\nu \mathcal{U}^\dag+\partial_\nu \mathcal{U}\partial_\mu \mathcal{U}^\dag)\right].
\end{align}

Moreover, the second term on the right-hand-side of Eq.~(\ref{gmU}) changes as
\begin{widetext}
\begin{align}\label{gmU2}
&\frac{1}{2}\text{Tr}\left(W'^\dag  W'\omega'_\mu\omega'_\nu+\omega'_\nu\omega'_\mu W'^\dag W'\right)\notag\\
=&\frac{1}{2}\text{Tr}\left[W^\dag  W(\omega_\mu\omega_\nu+\omega_\mu\partial_\nu \mathcal{U}\mathcal{U}^\dag+\partial_\mu \mathcal{U}\mathcal{U}^\dag\omega_\nu+\partial_\mu \mathcal{U}\mathcal{U}^\dag\partial_\nu \mathcal{U}\mathcal{U}^\dag)\right]+(\mu\leftrightarrow \nu)\notag\\
=&\frac{1}{2}\text{Tr}\left(W^\dag  W\omega_\mu\omega_\nu+\omega_\mu\omega_\nu W^\dag W\right)-\frac{1}{2}\text{Tr}\left[(W^\dagger W\omega_\mu+\omega_\mu W^\dagger W)\mathcal{U}\partial_\nu \mathcal{U}^\dag\right]
-\frac{1}{2}\text{Tr}\left[(W^\dagger W\omega_\nu+\omega_\nu W^\dagger W)\mathcal{U}\partial_\mu \mathcal{U}^\dag\right]\notag\\
-&\frac{1}{2}\text{Tr}\left[W^\dag W(\partial_\mu \mathcal{U}\partial_\nu \mathcal{U}^\dag+\partial_\nu \mathcal{U}\partial_\mu \mathcal{U}^\dag)\right],
\end{align}
where we have applied  $\partial_{\mu,\nu}\mathcal{U}\mathcal{U}^\dag=-\mathcal{U}\partial_{\mu,\nu}\mathcal{U}^\dag$, $\mathcal{U}^\dag\partial_\nu \mathcal{U}\mathcal{U}^\dag=-\partial_\nu \mathcal{U}^\dag$ and the cyclic property of the trace. Using Eqs.~(\ref{g2}) and (\ref{gmU2}), it is straightforward to verify that $g^{\text{U}\prime}_{\mu\nu}=g^\text{U}_{\mu\nu}$.

For the imaginary part $\sigma_{\mu\nu}$, the transformation (\ref{sigma1}) can be reformulated into
\begin{align}\label{Uf0}
\sigma'_{\mu\nu}=\sigma_{\mu\nu}+\frac{\mi}{2}\text{Tr}\left[\partial_\mu(W^\dag W\mathcal{U}\partial_\nu \mathcal{U}^\dag)-\partial_\nu(W^\dag W\mathcal{U}\partial_\mu \mathcal{U}^\dag)\right].
\end{align}
Similarly, the second term on the right-hand-side of Eq.~(\ref{Uf}) changes under $W'=W\mathcal{U}$ as
\begin{align}\label{Uf1}
&\frac{\mi}{2}\text{Tr}\left\{\partial_\mu\left[\mathcal{U}^\dag W^\dag  W\mathcal{U}(\mathcal{U}^\dag\omega_\nu \mathcal{U}+\mathcal{U}^\dag \partial_\nu \mathcal{U})\right]\right\}-(\mu\leftrightarrow \nu)\notag\\
=&\frac{\mi}{2}\text{Tr}\left[\partial_\mu(W^\dag W\omega_\nu)-\partial_\nu(W^\dag W\omega_\mu)\right]-\frac{\mi}{2}\text{Tr}\left[\partial_\mu(W^\dag W \mathcal{U}\partial_\nu\mathcal{U}^\dag)-\partial_\nu(W^\dag W \mathcal{U}\partial_\mu\mathcal{U}^\dag)\right]
\end{align}
Moreover, the second term in the last line cancels that of Eq.~(\ref{Uf0}), making $\sigma'^{\text{U}}_{\mu\nu}=\sigma^{\text{U}}_{\mu\nu}$.

\section{Details of Uhlmann metric, Bures metric, and Uhlmann form}\label{appd}
A calculation of Eq.~(\ref{gm3}) shows that $g^\text{U}_{\mu\nu}$ reduces to (see the proof below)
\begin{align}\label{gmUc}
g^\text{U}_{\mu\nu}=\text{Tr}\left(\partial_\mu \sqrt{\rho} \partial_\nu \sqrt{\rho}\right)+\frac{1}{2}\text{Tr}\left[ \rho( A_{\text{U}\mu}A_{\text{U}\nu}+A_{\text{U}\nu}A_{\text{U}\mu})\right].
\end{align}
Importantly, $g^\text{U}_{\mu\nu}$ is independent of the fiber $U$ when compared to its original expression (\ref{gm3}).
This resolves the paradox that $\dif s^2_\text{U}$ is a distance on $\mathcal{D}^N_N$ but seems to have an explicit dependence on the fiber $U$.
Taking the trace over the eigenstates of $\rho$, the first term of Eq.~(\ref{gmUc}) is
\begin{align}\label{gmUc2}
\text{Tr}\left(\partial_\mu \sqrt{\rho} \partial_\nu \sqrt{\rho}\right)=\sum_{ij}\langle i|\partial_\mu\sqrt{\rho}|j\rangle\langle j| \partial_\nu\sqrt{\rho}|i\rangle.
\end{align}
Using Eq.~(\ref{Umr1}) and interchanging the indices $i$ and $j$, the second term of Eq.~(\ref{gmUc}) becomes
$-\frac{1}{2}\sum_{ij}\frac{(\sqrt{\lambda_i}-\sqrt{\lambda_j})^2}{\lambda_i+\lambda_j}\langle i|\partial_\mu\sqrt{\rho}|j\rangle\langle j|\partial_\nu\sqrt{\rho}|i\rangle$.
Adding this to Eq.~(\ref{gmUc2}), we finally get the expression of the Uhlmann metric
\begin{align}\label{gmUf}
g^\text{U}_{\mu\nu}&=\sum_{ij}\left(1-\frac{1}{2}\frac{(\sqrt{\lambda_i}-\sqrt{\lambda_j})^2}{\lambda_i+\lambda_j}\right)\langle i|\partial_\mu\sqrt{\rho}|j\rangle\langle j|\partial_\nu\sqrt{\rho}|i\rangle\notag\\
&= g^\text{B}_{\mu\nu}
\end{align}
according to Eq.~(\ref{Bmb}).

To prove Eq.~(\ref{gmUc}), we use $W= \sqrt{\rho} U$, which leads to
\begin{align}\label{Ufc10}
\text{Tr}(\partial_\mu W^\dag \partial_\nu W)=\text{Tr}\left[\partial_\mu \sqrt{\rho} \partial_\nu \sqrt{\rho}+\rho\partial_\nu U\partial_\mu U^\dag+\sqrt{\rho}\partial_\nu\sqrt{\rho}U\partial_\mu U^\dag+\partial_\mu\sqrt{\rho}\sqrt{\rho}\partial_\nu U U^\dag\right].
\end{align}
Thus, the real part of the raw metric is
\begin{align}\label{Ufc1c}
\gamma_{\mu\nu}=\text{Tr}\left(\partial_\mu \sqrt{\rho} \partial_\nu \sqrt{\rho}\right)+\frac{1}{2}\text{Tr}\left[\rho(\partial_\mu U\partial_\nu U^\dag+\partial_\nu U\partial_\mu U^\dag)\right]+\frac{1}{2}\text{Tr}\left\{[\sqrt{\rho},\partial_\mu\sqrt{\rho}] U\partial_\nu U^\dag+[\sqrt{\rho},\partial_\nu\sqrt{\rho}] U\partial_\mu U^\dag\right\}.
\end{align}
Using the cyclic property of trace, we get the summation of the second and third terms of Eq.~(\ref{gm3})
\begin{align}\label{Ufc1e}
&\frac{1}{2}\text{Tr}\left[\rho (A_{\text{U}\mu}A_{\text{U}\nu}+A_{\text{U}\nu}A_{\text{U}\mu})-(\rho A_{\text{U}\mu}+A_{\text{U}\mu}\rho) U\partial_\nu U^\dag-(\rho A_{\text{U}\nu}+A_{\text{U}\nu}\rho) U\partial_\mu U^\dag -\rho (\partial_\mu U\partial_\nu U^\dag+\partial_\nu U\partial_\mu U^\dag) \right]\notag\\
=&\frac{1}{2}\text{Tr}\left[\rho (A_{\text{U}\mu}A_{\text{U}\nu}+A_{\text{U}\nu}A_{\text{U}\mu})-[\sqrt{\rho},\partial_\mu\sqrt{\rho}] U\partial_\nu U^\dag-[\sqrt{\rho},\partial_\nu\sqrt{\rho}]  U\partial_\mu U^\dag -\rho (\partial_\mu U\partial_\nu U^\dag+\partial_\nu U\partial_\mu U^\dag) \right]
\end{align}
The last three terms of Eq.~(\ref{Ufc1e}) exactly cancel the last three terms of Eq.~ (\ref{Ufc1c}), then the Uhlmann metric is given by Eq.~(\ref{gmUc}).

Using Eq.~(\ref{Ufc10}), the first term of Eq.~(\ref{Ufb}) is
\begin{align}\label{Ufc1}
\text{Tr}(\partial_\mu W^\dag \partial_\nu W)\dif R^\mu\wedge \dif R^\nu=\text{Tr}\left[\partial_\mu \sqrt{\rho} \partial_\nu \sqrt{\rho}+\rho\partial_\nu U\partial_\mu U^\dag+\sqrt{\rho}\partial_\nu\sqrt{\rho}U\partial_\mu U^\dag+\partial_\mu\sqrt{\rho}\sqrt{\rho}\partial_\nu U U^\dag\right]\dif R^\mu\wedge \dif R^\nu.
\end{align}
The first term vanishes since $\text{Tr}(\partial_\mu \sqrt{\rho} \partial_\nu \sqrt{\rho})$ is symmetric about $\mu$ and $\nu$. Interchanging $\mu$ and $\nu$ in the third term, we have
\begin{align}\label{Ufc2}
\text{Tr}(\partial_\mu W^\dag \partial_\nu W)\dif R^\mu\wedge \dif R^\nu&=\text{Tr}\left[\rho\partial_\nu U\partial_\mu U^\dag-\sqrt{\rho}\partial_\mu\sqrt{\rho}U\partial_\nu U^\dag-\partial_\mu\sqrt{\rho}\sqrt{\rho} U \partial_\nu U^\dag\right]\dif R^\mu\wedge \dif R^\nu\notag\\
&=\text{Tr}\left[\rho\partial_\nu U\partial_\mu U^\dag-\partial_\mu\rho U\partial_\nu U^\dag\right]\dif R^\mu\wedge \dif R^\nu.
\end{align}
Using $W^\dag W=U^\dag \rho U$ and the component form of Eq.~(\ref{AU4}), the second term of Eq.~(\ref{Ufb}) becomes
\begin{align}\label{Ufc3}
\text{Tr}\left\{\partial_\mu\left[U^\dag \rho U(U^\dag A_{\text{U}\nu} U+U^\dag\partial_\nu U)\right]\right\}\dif R^\mu\wedge \dif R^\nu
&=
\text{Tr}\left[\partial_\mu(\rho A_{\text{U}\nu})+\rho\partial_\nu U\partial_\mu U^\dag-\partial_\mu\rho U \partial_\nu U^\dag\right]\dif R^\mu\wedge \dif R^\nu.
\end{align}
Substituting Eqs.~(\ref{Ufc2}) and (\ref{Ufc3}) into Eq.~(\ref{Ufb}), we finally get Eq.~\eqref{Ufbf0}.

\section{Details of Sj$\ddot{\text{o}}$qvist distance}\label{app:Sdistance}
Here we briefly summarize the construction of the Sj$\ddot{\text{o}}$qvist distance. When a full-rank density matrix $\rho(t)$ evolves along a path, it is diagonalized as $\rho(t)=\sum_{n=0}^{N-1}\lambda_n(t)|n(t)\rangle\langle n(t)|$, from which one can define the spectral decomposition of $\rho(t)$ as $\{\sqrt{\lambda_n(t)}\me^{\mi\theta_n(t)}|n(t)\rangle\}_{n=0}^{N-1}$. Without loss of generality, let $\theta_n(0)=0$ and $\theta_n(t)$ be differentiable. For infinitesimal  $t$, the Sj$\ddot{\text{o}}$qvist distance between $\rho(t)$ and $\rho(0)=\rho$ is defined as
 \begin{align}\label{Sdis}
\dif^2_\text{S}(\rho(t),\rho)=\inf_{\theta_n(t)}\sum_n\big|\sqrt{\lambda_n(t)}\me^{\mi\theta_n(t)}|n(t)\rangle-\sqrt{\lambda_n}|n\rangle\big|^2
\end{align}
where $\lambda_n\equiv\lambda_n(0)$, $|n\rangle\equiv |n(0)\rangle$, and the infimum is taken among all possible sets of spectral phases $\{\theta_n(t)\}$. After some algebra, it is found that
 \begin{align}\label{Sdis2}
\dif^2_\text{S}(\rho( t),\rho)=2-2\sup_{\theta_n(t)}\sum_n\sqrt{\lambda_n\lambda_n(t)}|\langle n|n(t)\rangle|\cos \phi_n(t),
\end{align}
where $\cos \phi_n(t)=\dot{\theta}_n(t)t+\arg\left[1+\langle n(t)|\dot{n}(t)\rangle t\right]+O(t^2)$. Since $1+\langle n(t)|\dot{n}(t)\rangle t\approx \me^{\langle n(t)|\dot{n}(t)\rangle t}$, the infimum is obtained when
 \begin{align}\label{Sdispc}
\mi\dot{\theta}(t)+\langle n(t)|\dot{n}(t)\rangle=0, \quad \text{for} \quad n=0,\cdots, N-1.
\end{align}
Interestingly, this is precisely the parallel-transport condition associated with each individual pure state in the ensemble according to Eq. (\ref{BPen}).

With the help of purified states, the definition of the Sj$\ddot{\text{o}}$qvist distance can also be cast into the form similar to Eq. (\ref{db5}). The purifications of $\rho(t)$ is written as $W(t)=\sum_n\sqrt{\lambda_n(t)}|n(t)\rangle\langle n(t)|U(t)$. The key to the derivation is to choose the phase factor as $U(t)=\sum_n\me^{\mi\theta_n(t)}|n(t)\rangle\langle n|$, which is unitary due to the facts $U(t)U^\dag(t)=\sum_n|n(t)\rangle\langle n(t)|=1$ and $U^\dag(t)U(t)=\sum_n|n\rangle\langle n |=1$. Thus, the purification is $W(t)=\sum_n\sqrt{\lambda_n(t)}|n(t)\rangle\langle n|\me^{\mi\theta_n(t)}$, whose corresponding purified state is $|W(t)\rangle=\sum_n\sqrt{\lambda_n(t)}|n(t)\rangle\otimes\me^{\mi\theta_n(t)}|n\rangle$.

\section{Proof of Eq.~(\ref{Bdd0})}\label{appf}
Using $\sqrt{\rho}=\sum_j\sqrt{\lambda_j}|j\rangle\langle j|$, we have
 \begin{align}\label{dsr2}
 \left(\dif\sqrt{\rho}\right)^2&=\sum_j\left[(\dif\sqrt{\lambda_j})^2|j\rangle\langle j|+\sqrt{\lambda_j}\dif \sqrt{\lambda_j}(|\dif j\rangle\langle j|+|j\rangle\langle\dif j|)\right]+
 \sum_{jk}\sqrt{\lambda_k}\dif \sqrt{\lambda_j}(|j\rangle\langle j|\dif k\rangle\langle k|+| k\rangle\langle \dif k|j\rangle\langle j|)\notag\\
 &+\sum_j\lambda_j|\dif j\rangle\langle \dif j|+\sum_{jk}\sqrt{\lambda_k\lambda_j}(| \dif k\rangle\langle k| \dif  j\rangle\langle j|+| k\rangle\langle \dif k|j\rangle\langle \dif  j|+| k\rangle\langle \dif k|\dif  j\rangle\langle j|).
\end{align}
When taking trace with respect to the basis $\{|i\rangle\}$, the first term gives $\sum_i(\dif\sqrt{\lambda_i})^2$, the second and third terms vanish, the fourth term becomes
 \begin{align}\label{dsr3}
 \sum_{ij}\lambda_j\langle i|\dif j\rangle\langle \dif j|i\rangle=\sum_{j}\lambda_j\langle \dif j|\sum_i|i\rangle\langle i|\dif j\rangle=\sum_{j}\lambda_j\langle \dif j|\dif j\rangle,
\end{align}
and the last term gives
 \begin{align}\label{dsr4}
 \sum_{ij}\sqrt{\lambda_i\lambda_j}(\langle i|\dif j\rangle\langle  j|\dif i\rangle+\langle\dif i| j\rangle\langle \dif j| i\rangle)+\sum_i\lambda_i\langle \dif i|\dif i\rangle=2 \sum_{ij}\sqrt{\lambda_i\lambda_j}\langle i|\dif j\rangle\langle  j|\dif i\rangle+\sum_i\lambda_i\langle \dif i|\dif i\rangle.
\end{align}
Collecting all the results, we finally get Eq.~(\ref{Bdd0}).

\end{widetext}


\begin{thebibliography}{83}
\bibitem [{Ref()}]{Referee}%
	\BibitemOpen
	\href@noop {} {}\bibinfo {note} {The symmetry was pointed out by an anonymous
		referee.}\BibitemShut {Stop}%
\end{thebibliography}
%

\end{document}